\documentclass[pra,preprint,notitlepage,nofootinbib,a4paper]{revtex4}

\usepackage{graphicx}
\usepackage{bm,amssymb,amsmath,textcomp}
\usepackage{array,dcolumn}
\usepackage%[colorlinks]  
{hyperref}

\def\la{\;
\raise0.3ex\hbox{$<$\kern-0.75em\raise-1.1ex\hbox{$\sim$}}\; }
\def\ga{\;
\raise0.3ex\hbox{$>$\kern-0.75em\raise-1.1ex\hbox{$\sim$}}\; }

\newcommand{\dmm}{$\Delta\mu/\mu$}
\newcommand{\daa}{$\Delta\alpha/\alpha$}
\newcommand{\kms}{km~s$^{-1}$}

\newcommand{\cmo}{cm$^{-1}$}

\newcommand{\fref}[1]{Fig.~\ref{#1}}
\newcommand{\sref}[1]{Sec. \ref{#1}}
\newcommand{\Eref}[1]{Eq.~(\ref{#1})}
\newcommand{\tref}[1]{Table~\ref{#1}}
\newcommand{\rtw}{\longrightarrow}

\begin{document}

\title{Microwave and submillimeter molecular transitions and their dependence
on fundamental constants}

\author{M. G. Kozlov$^{1,2}$}
\author{S. A. Levshakov$^{2,3}$}

\affiliation{$^1$Petersburg Nuclear Physics Institute, 188300 Gatchina}

\affiliation{$^2$St.~Petersburg Electrotechnical University ``LETI'', Prof.
Popov Str. 5, 197376 St.~Petersburg}

\affiliation{$^3$Ioffe Physical-Technical Institute, Polytekhnicheskaya Str.
26, 194021 St.~Petersburg}

\begin{abstract}
Microwave and submillimeter molecular transition frequencies between nearly
degenerated rotational levels, tunneling transitions, and mixed
tunneling-rotational transitions show an extremely high sensitivity to the
values of the fine-structure constant, $\alpha$, and the electron-to-proton
mass ratio, $\mu$. This review summarizes the theoretical background on
quantum-mechanical calculations of the sensitivity coefficients of such
transitions to tiny changes in $\alpha$ and $\mu$ for a number of molecules
which are usually observed in Galactic and extragalactic sources,
and discusses the possibility of testing the space- and time-invariance of
fundamental constants through comparison between precise laboratory
measurements of the molecular rest frequencies and their astronomical
counterparts. In particular, diatomic radicals CH, OH, NH$^+$, and
a linear polyatomic radical C$_3$H in $\Pi$ electronic ground state,
polyatomic molecules NH$_3$, ND$_3$, NH$_2$D, NHD$_2$, H$_2$O$_2$, H$_3$O$^+$,
CH$_3$OH, and CH$_3$NH$_2$ in their tunneling and tunneling-rotational modes are considered.
It is shown that sensitivity coefficients strongly depend on the quantum numbers
of the corresponding transitions. This can be used for astrophysical tests of
Einstein's Equivalence Principle all over the Universe at an unprecedented
level of sensitivity of $\sim 10^{-9}$, which is a limit three to two orders
of magnitude lower as compared to the current constraints on cosmological
variations of $\alpha$ and $\mu$: \daa\ $< 10^{-6}$, \dmm\ $< 10^{-7}$.
\end{abstract}

 \maketitle

\newpage

\newpage

\section{Introduction}
\label{sect-1}

The fundamental laws of particle  physics, in our current understanding,
depend on 28 constants including the gravitational constant, $G$, the mass,
$m_{\rm e}$, and charge, $e$, of the electron, the masses of six quarks,
$m_{\rm u}$, $m_{\rm d}$, $m_{\rm c}$, $m_{\rm s}$, $m_{\rm t}$, and $m_{\rm
b}$, the Planck constant, $\hbar$, the Sommerfeld constant $\alpha$, the
coupling constants of the weak, $g_{\rm w}$, and strong, $g_{\rm s}$,
interactions, etc. The numerical values of these constants are not calculated
within the Standard Model and remain, as Feynman wrote about the fine
structure constant $\alpha$ in 1985, ``one of the greatest mysteries of
physics''~\cite{29}. However, it is natural to ask whether these
constants are really constants, or whether they vary with the age of the
universe, or over astronomical distances.

The idea that the fundamental constants may vary on the cosmological time
scale has been discussing in different forms since 1937, when Milne and Dirac
argued about possible variations of the Newton constant $G$ during the
lifetime of the universe~\cite{72,23}. Over the past few decades,
there have been extensive searches for persuasive evidences of the variation of
physical constants. So far, there was found no one of them. The current limits
for dimensionless constants such as the fine structure constant, 
$\alpha = e^2/\hbar c$, and the electron to proton mass ratio, 
$\mu = m_{\rm e}/m_{\rm p}$, obtained in laboratory experiments 
and from the Oklo natural reactor are on the order of one part in 
$10^{15}-10^{17}$~\cite{17,83,90} 
and one part in $10^{14}-10^{16}$~\cite{93,10,28}
per year, respectively. The detailed discussion of ideas 
behind laboratory experiments can be found in a review~\cite{30}. 

Assuming that the constants are linearly dependent on the cosmic time, the
same order of magnitude constraints on the fractional changes in 
$\Delta \alpha/\alpha = (\alpha_{\rm obs} - \alpha_{\rm lab})/\alpha_{\rm lab}$ and in
$\Delta \mu/\mu = (\mu_{\rm obs} - \mu_{\rm lab})/\mu_{\rm lab}$ are stemming
from astronomical observations of extragalactic objects at redshifts $z
\sim 1-5$~\cite{73,2,46,59,58}.
Less stringent constraints at a
percent level have been obtained from the cosmic microwave background (CMB) at
$z \sim 10^3$~\cite{57,69,78} 
and big bang nucleosynthesis (BBN) at $z \sim 10^{10}$~\cite{31,18}. 
We note that space and/or time dependence of
$\alpha$ based on optical spectra of quasars and discussed in the literature~\cite[and references therein]{99}
is still controversial and
probably caused by systematic effects since independent radio-astronomical
observations, which are more sensitive, show only null results for both
$\Delta \alpha/\alpha$ and $\Delta\mu/\mu$~\cite{87,4}. 

Surprisingly, it looks as if the Einstein heuristic principle of local
position invariance (LPI)~--- {\it the outcome of any local non-gravitational
experiment is independent of where and when in the universe it is
performed}~--- is valid all over the universe, i.e., at the level of
$\sim10^{-6}$ neither $\alpha$ no $\mu$ deviate from their terrestrial values
for the passed $10^{10}$ yr. In the Milky Way, it was also found no
statistically significant deviations of \dmm\ from zero at even more deeper
level of $\sim10^{-8}$~\cite{61,60,27}. 

However, the violation of the LPI was predicted in some theoretical models
such as, for example, the theory of superstrings which considers time
variations of $\alpha$, $g_{\rm w}$, and the QCD scale $\Lambda_{\rm QCD}$
(i.e., $\mu$ since $m_{\rm p} \propto \Lambda_{\rm QCD}$) and thereby opening
a new window on physics beyond the Standard Model~\cite[and references therein]{19}. 
If the fundamental constants are found to be changing in
space and time, then they are not absolute but dynamical quantities which
follow some deeper physical laws that have to be understood.
Already present upper limits on the variation of the fundamental
constants put very strong constraints on the theories beyond the Standard
Model~\cite[and references therein]{97}. 
This motivates
the need for more precise laboratory and astronomical tests of the LPI. 
Of course, there are also other attempts to look for the new physics. For
example the electric dipole moments (EDMs) of the elementary particles are
very sensitive to the different extensions of the Standard Model. Present
limit on the EDM of the electron significantly constrains supersymmetrical
models and other theories~\cite{88,34}. 

In this review we will consider tests of LPI which are based on the analysis
of microwave and submillimeter\footnote{The frequency range 1~GHz $\leq \nu
\leq 300$~GHz is usually referred to as a microwave range. Molecular
transitions below 1~GHz (wavelength $\lambda > 30$ cm) are from a
low-frequency range which is restricted by the ionospheric cut-off at 10~MHz
($\lambda = 30$ m).} astronomical spectra and which are essentially more
sensitive to small variations in $\alpha$ and $\mu$ than the test based on
optical spectral observations of quasars.

\section{Differential measurements of \daa\ and \dmm\ from atomic and molecular
spectra of cosmic objects}
\label{sect-2}

Speaking about stable matter, as, for example, atoms and molecules, we have
only seven physical constants that describe their spectra~\cite{33}: 
$$
G, \Lambda_{\rm QCD}, \alpha, m_{\rm e}, m_{\rm u}, m_{\rm d}, m_{\rm s} \, .
$$
The QCD scale parameter $\Lambda_{\rm QCD}$ and the masses of the light quarks
u, d, and s contribute to the nucleon mass $m_{\rm p}$ (with 
$\Lambda_{\rm QCD} \gg m_{\rm u}+m_{\rm d}+m_{\rm s}$) and, 
thus, the electron-to-proton mass ratio $\mu$ is a physical constant characterizing 
the strength of electroweak interaction in terms of the strong interaction.

In the nonrelativistic limit and for an infinitely heavy pointlike nucleus all
atomic transition frequencies are proportional to the Rydberg constant, $R$,
and the ratios of atomic frequencies do not depend on any fundamental
constants. Relativistic effects cause corrections to atomic energy, which can
be expanded in powers of $\alpha^2$ and $\alpha^2Z^2$, the leading term being
$\alpha^2Z^2R$, where $Z$ is atomic number.  Corrections accounting for the
finite nuclear mass are proportional to $R\mu/Z$, but for atoms they are much
smaller than relativistic corrections.

Astronomical differential measurements of the dimensionless constants $\alpha$
and $\mu$ are based on the comparison of the line centers in the
absorption/emission spectra of cosmic objects and the corresponding laboratory
values. It follows that the uncertainties of the laboratory rest frequencies
and the line centers in astronomical spectra are the prime concern of such
measurements. It is easy to estimate the natural bounds set by these
uncertainties on the values of \daa\ and \dmm.

Consider the dependence of an atomic frequency $\omega$ on $\alpha$ in the
comoving reference frame of a distant object located at redshift $z$~\cite{24,25}:
\begin{equation}
\omega_z = \omega + qx + O(x^2), \,\,\, x \equiv (\alpha_z/\alpha)^2 - 1.
\label{Sect2Eq1}
\end{equation}
Here $\omega$ and $\omega_z$ are the frequencies corresponding to the
present-day value of $\alpha$ and to a change $\alpha \rightarrow \alpha_z$ at
a redshift $z$. In this relation, the so-called $q$ factor is an individual
parameter for each atomic transition.

If $\alpha_z \neq \alpha$, the quantity $x$ in (\ref{Sect2Eq1}) differs from
zero and the corresponding frequency shift $\Delta\omega = \omega_z - \omega$
is given by
\begin{equation}
\frac{\Delta\omega}{\omega} = Q\frac{\Delta\alpha}{\alpha}\, ,
\label{Sect2Eq2}
\end{equation}
where $Q = 2q/\omega$ is the dimensionless sensitivity coefficient and
$\Delta\alpha = (\alpha_z - \alpha)/\alpha$ is the fractional change in
$\alpha$. Here we assume that $|\Delta\alpha/\alpha| \ll 1$. The condition
$\alpha_z \neq \alpha$ leads to a change in the apparent redshift of the
distant object $\Delta z = \tilde{z} - z$:
\begin{equation}
\frac{\Delta\omega}{\omega} = -\frac{\Delta z}{1+z} \equiv 
\frac{\Delta v}{c}\ , 
\label{Sect2Eq3}
\end{equation}
where $\Delta v$ is the Doppler radial velocity shift.

If $\omega'$ is the observed frequency from the distant object, then the true
redshift is given by
\begin{equation}
1 + z = \frac{\omega_z}{\omega'}\, , 
\label{Sect2Eq4}
\end{equation}
whereas the shifted (apparent) value is
\begin{equation}
1 + \tilde{z} = \frac{\omega}{\omega'}\, . 
\label{Sect2Eq5}
\end{equation}
Now, if we have two lines of the same element with the apparent redshifts
$\tilde{z}_1$ and $\tilde{z}_2$ and the corresponding sensitivity coefficients
$Q_1$ and $Q_2$, then
\begin{equation}
\Delta Q \frac{\Delta\alpha}{\alpha} = \frac{\tilde{z}_1 - \tilde{z}_2}{1 + z}
= \frac{\Delta v}{c} \ . 
\label{Sect2Eq6}
\end{equation}
Here $\Delta v = v_1 - v_2$ is the difference of the measured radial
velocities of these lines, and $\Delta Q = Q_2 - Q_1$ is the corresponding
difference between their sensitivity coefficients. By comparing the apparent
redshifts of two lines with different sensitivity coefficients $Q$ we can
study variation of $\alpha$ on a cosmological timescale.

Unfortunately, optical and UV transitions of atoms and molecules are not very
sensitive to changes in $\alpha$ and $\mu$. The sensitivity coefficients of
atomic resonance transitions of usually observed in quasar spectra chemical elements 
(C, N, O, Na, Mg, Al, Si, S, Ca, Ti, Cr, Mn, Fe, Co, Ni, Zn) are very small, $Q
\sim (\alpha Z)^2 \ll 1$~\cite{6}. 
The same order of magnitude sensitivity coefficients to $\mu$ variations have been
calculated for the UV transitions in the Lyman and Werner bands of molecular
hydrogen H$_2$~\cite{98,71,96}, 
and for the UV transitions in the 4{$th$} positive band system
$A^1\Pi - X^1\Sigma^+$ of carbon monoxide CO~\cite{91}.

Small values of $Q$ and $\Delta Q$ put tough constraints on optical methods to
probe \daa\ and \dmm. Let us consider an example of Fe\,{\sc ii} lines arising
from the ground state $3d^6(^5D)4s$. In quasar spectra we observe 7 resonance
transitions ranging from 1608 \AA\ to 2600 \AA\ with both signs sensitivity
coefficients: $Q_{\lambda1608} = -0.0322,\  Q_{\lambda1611} = +0.0502,$ and $Q
\simeq +0.08$ for transitions with $\lambda > 2000$ \AA\ \cite[note a factor of two 
difference in the definition of the coefficients $Q$ with the present work]{85}. 
This gives us the maximum value of $\Delta Q \simeq 0.11$
which is known with an error of $\sim 30$\%. From (\ref{Sect2Eq6}) it follows
that a variance of $\Delta\alpha/\alpha \sim 10^{-5}$ would induce a velocity
offset $\Delta v \simeq 0.3$ \kms\ between the 1608 \AA\ line and any of the
line with $\lambda > 2000$ \AA. We may neglect uncertainties of the rest frame 
wavelengths since they are $\sim 0.02$ \kms\ \cite{79}.
If both iron line centers are measured in quasar spectra with the same error
$\sigma_v$, then the error of the offset $\Delta v$ is $\sigma_{\Delta v} =
\sqrt{2}\sigma_v$. The error $\sigma_{\Delta v}$ is a statistical estimate of
the uncertainty of $\Delta v$, and, hence, it should be less than the absolute
value of $\Delta v$. This gives us the following inequality to adjust
parameters of spectral observations required to probe \daa\ at a given level:
\begin{equation}
\sigma_v < \frac{\Delta Q}{\sqrt{2}} \frac{\Delta \alpha}{\alpha} c \ .
\label{Sect2Eq7}
\end{equation}
At  $\Delta \alpha/\alpha \sim 10^{-5}$, the required position accuracy should
be $\sigma_v \la 0.25$ \kms. A typical error of the line center of an
unsaturated absorption line in quasar spectra is about 1/10$th$ of the pixel
size (the wavelength interval between pixels)~\cite{63}. 
Current observations with the UV-Visual Echelle Spectrograph (UVES) at the ESO
Very Large Telescope (VLT) provide a pixel size $\Delta \lambda_{\rm pix} \sim
0.05-0.06$ \AA, i.e., at $\lambda \sim 5000$ \AA\ the expected error
$\sigma_v$ should be $\sim 0.3$ \kms, which is comparable to the velocity
offset due to a fractional change in $\alpha$ at the level of $10^{-5}$. Such
a critical relationship between the `signal' (expected velocity offset $\Delta
v$) and the error $\sigma_v$ hampers measuring $\Delta\alpha/\alpha$ at the
level of $ \sim 10^{-5}$ from any absorption system taking into account all
imperfections of the spectrograph and the data reduction procedure. Systematic
errors exceeding 0.5 \kms\ are known to be typical for the wavelength
calibration in both the VLT/UVES and Keck/HIRES spectrographs~\cite{2,35,101,1}. 
At this level of the systematic errors an estimate of \daa\ from any individual
absorption-line system must be considered as an {\it upper limit} but not a
`signal'. Otherwise, a formal statistical analysis of such values may lead to
unphysical results (examples can be found in the literature).

The UV molecular spectra of H$_2$ and CO observed at high redshifts in the
optical wavelength band encounter with similar difficulties and restrictions.
The maximum difference between the sensitivity coefficients in case of H$_2$
is $\Delta Q \sim 0.06$, the rest frame wavelength uncertainties are
negligible, $\sim 5\times10^{-9}$~\cite{92},  
and with the current spectral facilities at giant
telescopes it is hard to get estimates of \dmm\ at a level deeper than
$10^{-5}$. For carbon monoxide such measurements have not been done so far but
the expected limit on \dmm\ should be $\ga 10^{-5}$ since CO lines are much
weaker than H$_2$~\cite{80} and therefore their line
centers are less certain. The analogue of Eq.(\ref{Sect2Eq6}) for the
$\mu$-estimation from a pair of molecular lines is~\cite{64}: 
\begin{equation}
\frac{\Delta \mu}{\mu} = \frac{\Delta v}{c \Delta Q} = \frac{v_1 - v_2}{c (Q_2
- Q_1)} \ , 
\label{Sect2Eq8}
\end{equation}
and for a given level of \dmm, molecular line centers should be measured with
an error
\begin{equation}
\sigma_v < \frac{\Delta Q}{\sqrt{2}} \frac{\Delta \mu}{\mu} c \ .
\label{Sect2Eq9}
\end{equation}
This means that at  $\Delta \mu/\mu \sim 10^{-5}$, the required position
accuracy should be $\sigma_v \la 0.13$ \kms, or the pixel size $\Delta
\lambda_{\rm pix} \la 0.017$ \AA\ at 4000 \AA. This requirement was realized
in the VLT/UVES observations of the quasar Q0347--383~\cite{100} where
a limit on \dmm\ of $(4.3\pm7.2)\times10^{-6}$ was set.

At present the only way to probe variation of the fundamental constants on the
cosmological timescale at a level deeper than $10^{-5}$ is to switch from
optical to far infrared and microwave bands. In the microwave, or
submillimeter range there are a good deal of molecular transitions arising in
Galactic and extragalactic sources. Electronic, vibrational, and rotational
energies in molecular spectra are scaled as $E_{\rm el} : E_{\rm vib} : E_{\rm
rot} = 1:\mu^{1/2}:\mu$. In other words, the sensitivity coefficients for pure
vibrational and rotational transitions are equal to $Q_\mu=0.5$ and $Q_\mu=1$,
respectively. Besides, molecules have fine and hyperfine structures,
$\Lambda$-doubling, hindered rotation, accidental degeneracy between narrow
close-lying levels of different types, which have a specific dependence on the
physical constants. The advantage of radio observations is that some of these
molecular transitions are approximately 100-1000 times more sensitive to
variations of $\mu$ and/or $\alpha$ than optical and UV transitions.

In the far infrared waveband also lie atomic fine-structure transitions,
which have sensitivity to $\alpha$-variation $Q_\alpha\approx 2$~\cite{55}. 
We can combine observations of these lines and rotational molecular
transitions to probe a combination $F = \alpha^2/\mu$~\cite{62}. 
Besides, radio-astronomical observations allow us to measure emission lines
from molecular clouds in the Milky Way with an extremely high spectral
resolution (channel width $\sim 0.02$ \kms) leading to stringent constraints
at the level of $\sim 10^{-9}$~\cite{61}. The level $10^{-9}$ is
a natural limit for radio-astronomical observations since it requires the rest
frequencies of molecular transitions to be known with an accuracy better than
100 Hz. At the moment only ammonia inversion transitions and 18 cm OH
$\Lambda$-doublet transitions have been measured in the laboratory with such a
high accuracy~\cite{56,41}. 

In the next sections we consider in more detail the sensitivities of different
types of molecular transitions to changes in $\alpha$ and $\mu$. We are mainly
dealing with molecular lines observed in microwave and submillimeter ranges in
the interstellar medium, but a few low-frequency transitions with high
sensitivities are also included in our analysis just to extend the list of
possible targets for future studies at the next generation of large telescopes
for low-frequency radio astronomy.

\section{Diatomic radicals in the $\Pi$ ground state: CH, OH, and NH$^+$}
\label{sect-3}

We start our analysis of the microwave spectra of molecules from the simplest
systems --- diatomic molecules with nonzero projection of the electronic
angular momentum $\bm L$ on the molecular axis. Several such molecules are
observed in the interstellar medium. Here we will mostly focus on the two most
abundant species --- CH and OH. Recently it was realized that
$\Lambda$-doublet transitions in these molecules have high sensitivity to the
variation of both $\alpha$ and $\mu$~\cite{15,20,51}. 
There are also several relatively low frequency transitions
between rotational levels of the ground state doublet $\Pi_{1/2}$ and
$\Pi_{3/2}$ with sensitivities, which are significantly different from the
typical rotational ones~\cite{22}. Then we will briefly discuss
the NH$^+$ radical\footnote{NH$^+$ has not yet been detected in space, its
fractional abundance in star-forming regions is estimated
$N$(NH$^+$)/$N$(H$_2$) $\la 4\times10^{-10}$~\cite{82}.},
which is interesting because it has very low lying excited
electronic state $^4\Sigma^-$. This leads to an additional enhancement of the
dimensionless sensitivity coefficients $Q$~\cite{5}. The latter are
defined as follows:
 \begin{align}
 \frac{\Delta\omega}{\omega}
 = Q_\alpha\frac{\Delta\alpha}{\alpha}
 + Q_\mu\frac{\Delta\mu}{\mu}\,.
\label{Q-factors}
 \end{align}

\subsection{$\Lambda$-doubling and $\Omega$-doubling}
\label{analytic}

Consider electronic state with nonzero projection $\Lambda$ of the orbital
angular momentum on the molecular axis. The spin-orbit interaction couples
electron spin $\bm{S}$ to the molecular axis, its projection being $\Sigma$.
To a first approximation the spin-orbit interaction is reduced to the form
$H_{so}=A\Lambda\Sigma$. Total electronic angular momentum
$\bm{J}_e=\bm{L}+\bm{S}$ has projection $\Omega$ on the axis,
$\Omega=\Lambda+\Sigma$. For a particular case of $\Lambda=1$ and $S=\tfrac12$
we have two states $\Pi_{1/2}$ and $\Pi_{3/2}$ and the energy difference
between them is: $E(\Pi_{3/2})-E(\Pi_{1/2})=A$.

Rotational energy of the molecule is described by the Hamiltonian:
 \begin{subequations}\label{Hrot}
 \begin{align}
 \label{Hrot1}
  H_\mathrm{rot} &=B(\bm{J}-\bm{J}_e)^2\\
  \label{Hrot2}
  &=B\bm{J}^2-2B(\bm{JJ}_e)+B\bm{J}_e^2\,,
 \end{align}
 \end{subequations}
where $B$ is the rotational constant and $\bm{J}$ is the total angular
momentum of the molecule. The first term in expression \eqref{Hrot2} describes
conventional rotational spectrum. The last term is constant for a given
electronic state and can be added to the electronic energy.\footnote{Note that
this term contributes to the separation between the states $\Pi_{1/2}$ and
$\Pi_{3/2}$. This becomes particularly important for light molecules, where
the constant $A$ is small.} The second term describes $\Omega$-doubling and is
known as the Coriolis interaction $H_\mathrm{Cor}$.

If we neglect the Coriolis interaction, the eigenvectors of Hamiltonian
\eqref{Hrot} have definite projections $M$ and $\Omega$ of the molecular
angular momentum $\bm{J}$ on the laboratory axis and on the molecular axis
respectively. In this approximation the states
$|J,M,\Lambda,\Sigma,\Omega\rangle$ and $|J,M,-\Lambda,-\Sigma,-\Omega\rangle$
are degenerate, $E_{J,\pm\Omega}=BJ(J+1)$. The Coriolis interaction couples
these states and removes degeneracy. New eigenstates are the states of
definite parity $p=\pm 1$~\cite{11}:
 \begin{align}\label{patity_states}
 |J,M,\Omega,p\rangle
 &= \left(|J,M,\Omega\rangle
 +p(-1)^{J-S} |J,M,-\Omega\rangle\right)/\sqrt{2}\, .
 \end{align}
The operator $H_\mathrm{Cor}$ can only change quantum number $\Omega$ by one,
so the coupling of states $|\Omega\rangle$ and $|-\Omega\rangle$ takes place
in the $2\Omega$ order of the perturbation theory in $H_\mathrm{Cor}$.

The $\Omega$-doubling for the state $\Pi_{1/2}$ happens already in the first
order in the Coriolis interaction, but has additional smallness from the
spin-orbit mixing. The operator $H_\mathrm{Cor}$ can not directly mix
degenerate $|\Lambda,\Sigma,\Omega\rangle$ states
$|1,-\tfrac12,\tfrac12\rangle$ and $|-1,\tfrac12,-\tfrac12\rangle$ because it
requires changing $\Lambda$ by two. Therefore, we need to consider spin-orbit
mixing of the $\Pi$ and $\Sigma$ states:
 \begin{align}\label{so-mixing}
 |\Omega=\tfrac12\rangle
 &=|1,-\tfrac12,\tfrac12\rangle
 +\zeta
 |0,\tfrac12,\tfrac12\rangle,
 \end{align}
where
 \begin{align}\label{so-mixing1}
 \zeta\sim A/(E_\Pi-E_\Sigma),
 \end{align}
 and then
 \begin{align}\label{Pi_1/2}
 \langle\Omega=\tfrac12|H_\mathrm{Cor}|\Omega=-\tfrac12\rangle
 =2\zeta B (J+\tfrac12) \langle\Lambda=1|L_x|\Lambda=0\rangle.
 \end{align}
Note that $\zeta$ depends on the non-diagonal matrix element (ME) of the
spin-orbit interaction and \Eref{so-mixing1} is only an order of magnitude
estimate. It is important, though, that non-diagonal and diagonal MEs have
similar dependence on fundamental constants. We conclude that
$\Omega$-splitting for the $\Pi_{1/2}$ level must scale as
$ABJ/(E_\Pi-E_\Sigma)$. The $\Omega$-doubling for $\Pi_{3/2}$ state takes
place in the third order in the Coriolis interaction. Here $H_\mathrm{Cor}$
has to mix first states $\Pi_{3/2}$ with $\Pi_{1/2}$ and $\Pi_{-3/2}$ with
$\Pi_{-1/2}$ before ME \eqref{Pi_1/2} can be used. Therefore, the splitting
scales as $B^3J^3/[A(E_\Pi-E_\Sigma)]$.

The above consideration corresponds to the coupling case $a$, when $|A|\gg B$.
In the opposite limit the states $\Pi_{1/2}$ and $\Pi_{3/2}$ are strongly
mixed by the Coriolis interaction and spin $\bm{S}$ decouples from the
molecular axis (coupling case $b$). As a result, the quantum numbers $\Sigma$
and $\Omega$ are not defined and we only have one quantum number $\Lambda=\pm 1$.
The $\Lambda$-splitting takes place now in the second order in the
Coriolis interaction via intermediate $\Sigma$ states. The scaling here is
obviously of the form $B^2J^2/(E_\Pi-E_\Sigma)$. Note that in contrast to the
previous case $|A|\gg B$, the splitting here is independent on $A$.

We can now use found scalings of the $\Lambda$- and $\Omega$-doublings to
determine sensitivity coefficients \eqref{Q-factors}. We only need to recall
that in atomic units $A\propto\alpha^2$ and $B\propto\mu$. We conclude that
for the case $a$ the $\Omega$-doubling spectrum has following sensitivity
coefficients:
 \begin{subequations}\label{doubling}
 \begin{align}
 \label{doubling1}
 &\mathrm{State\,}^2\Pi_{1/2}:
 \quad Q_\alpha = \phantom{-}2\,,
 \quad Q_\mu = 1\,,\\
 \label{doubling2}
 &\mathrm{State\,}^2\Pi_{3/2}:
 \quad Q_\alpha = -2\,,
 \quad Q_\mu = 3\,.
 \end{align}
For the case $b$, when $\bm{S}$ is completely decoupled from the
axis, the $\Lambda$-doubling spectrum has following sensitivity
coefficients:
 \begin{align}
 \label{doubling3}
 &\mathrm{State\,\,}\Pi:
 \quad Q_\alpha = 0\,,
 \quad Q_\mu = 2\,.
 \end{align}
 \end{subequations}

When constant $A$ is slightly larger than $B$, the spin $\bm{S}$ is coupled to
the axis only for lower rotational levels. As rotational energy grows with $J$
and becomes larger than the splitting between states $\Pi_{1/2}$ and
$\Pi_{3/2}$, the spin decouples from the axis. Consequently, the
$\Omega$-doubling is transformed into $\Lambda$-doubling. Equations
\eqref{doubling} show that this can cause significant changes in sensitivity
coefficients. The spin-orbit constant $A$ can be either positive (CH
molecule), or negative (OH). The sign of the $\Omega$-doubling depends on the
sign of $A$, while $\Lambda$-doubling does not depend on $A$ at all.
Therefore, decoupling of the spin can change the sign of the splitting. In
\sref{numeric} we will see that this can lead to a dramatic enhancement of the
sensitivity to the variation of fundamental constants.

%--------------------------------Table I
\begin{table*}[tbh]
\caption{Frequencies (in MHz) and sensitivity coefficients for hyperfine
components $(J,F\rightarrow J,F')$ of $\Lambda$-doublet lines in CH and OH
molecules. Recommended frequencies and their uncertainties are taken from~\cite{66,84,77}. }
\label{tab-1}
\begin{tabular}{lldccdcdddd}
 \hline\hline\\[-7pt]
 \multicolumn{1}{c}{Molecule}
 &\multicolumn{1}{c}{Level}
 &\multicolumn{1}{c}{$J$} &\multicolumn{1}{c}{$F$}
 &\multicolumn{1}{c}{$F'$} &\multicolumn{4}{c}{$\omega$~(MHz)}
 &\multicolumn{1}{c}{$Q_\alpha$}
 &\multicolumn{1}{c}{$Q_\mu$}
\\
 \cline{6-9}\\[-7pt]
&&&&&\multicolumn{1}{c}{Recom.} &\multicolumn{1}{c}{Uncert.}
&\multicolumn{1}{c}{Theory} &\multicolumn{1}{c}{Diff.}
\\
\hline
 $^{12}$CH & $^2\Pi_{1/2}$
 & 0.5 & 0 & 1 &   3263.795&   0.003 & 3269.40 &  -5.61 &  0.59  &   1.71   \\
&& 0.5 & 1 & 1 &   3335.481&   0.001 & 3340.77 &  -5.29 &  0.62  &   1.70   \\
&& 0.5 & 1 & 0 &   3349.194&   0.003 & 3354.11 &  -4.92 &  0.63  &   1.69   \\[1mm]
&& 1.5 & 1 & 2 &   7275.004&   0.001 & 7262.25 &  12.75 & -0.24  &   2.12   \\
&& 1.5 & 1 & 1 &   7325.203&   0.001 & 7312.02 &  13.18 & -0.23  &   2.11   \\
&& 1.5 & 2 & 2 &   7348.419&   0.001 & 7335.30 &  13.12 & -0.22  &   2.11   \\
&& 1.5 & 2 & 1 &   7398.618&   0.001 & 7385.08 &  13.54 & -0.20  &   2.10   \\[1mm]
 $^{12}$CH & $^2\Pi_{3/2}$
 & 1.5 & 2 & 2 &    701.68 &   0.01  &  682.96 &  18.72 & -8.44  &   6.15   \\
&& 1.5 & 1 & 2 &    703.97 &   0.03  &  679.83 &  24.14 & -8.66  &   6.32   \\
&& 1.5 & 2 & 1 &    722.30 &   0.03  &  702.98 &  19.52 & -8.37  &   6.17   \\
&& 1.5 & 1 & 1 &    724.79 &   0.01  &  699.85 &  24.94 & -8.07  &   5.97   \\[1mm]
 $^{16}$OH & $^2\Pi_{3/2}$
 & 1.5 & 1 & 2 &   1612.2310& 0.0002 & 1595.42 &  16.81 &  -1.27 &   2.61   \\
&& 1.5 & 1 & 1 &   1665.4018& 0.0002 & 1648.93 &  16.47 &  -1.14 &   2.55   \\
&& 1.5 & 2 & 2 &   1667.3590& 0.0002 & 1650.66 &  16.70 &  -1.14 &   2.55   \\
&& 1.5 & 2 & 1 &   1720.5300& 0.0002 & 1704.17 &  16.36 &  -1.02 &   2.49   \\[1mm]
 $^{16}$OH & $^2\Pi_{1/2}$
 & 0.5 & 0 & 1 &   4660.2420& 0.0030 & 4638.98 &  21.26 &   2.98 &   0.50   \\
&& 0.5 & 1 & 1 &   4750.6560& 0.0030 & 4729.51 &  21.15 &   2.96 &   0.51   \\
&& 0.5 & 1 & 0 &   4765.5620& 0.0030 & 4744.50 &  21.06 &   2.96 &   0.51   \\[1mm]
&& 4.5 & 5 & 4 &     88.9504& 0.0011 &   64.34 &  24.61 &-921.58 & 459.86   \\
&& 4.5 & 5 & 5 &    117.1495& 0.0011 &   92.35 &  24.80 &-699.65 & 349.59   \\
&& 4.5 & 4 & 4 &    164.7960& 0.0011 &  141.20 &  23.60 &-496.67 & 248.77   \\
&& 4.5 & 4 & 5 &    192.9957& 0.0011 &  169.22 &  23.78 &-424.05 & 212.68   \\
\hline\hline
\end{tabular}
\end{table*}

\subsection{Intermediate coupling}
\label{numeric}

The $\Lambda$-doubling for the intermediate coupling was studied in detail in
many papers, including~\cite{70,13,12} (see also the book~\cite{11}). 
Here we use the effective Hamiltonian $H_\mathrm{eff}$ from~\cite{70} 
in the subspace of the levels $\Pi_{1/2}^\pm$ and $\Pi_{3/2}^\pm$, where upper
sign corresponds to the parity $p$ in \Eref{patity_states}. The operator
$H_\mathrm{eff}$ includes spin-rotational and hyperfine parts
 \begin{align}\label{Heff1}
 H_\mathrm{eff} &=H_\mathrm{sr} + H_\mathrm{hf}\,.
 \end{align}
Neglecting third order terms in the Coriolis and spin-orbit interactions, we
get the following simplified form of the spin-rotational part:
 \begin{subequations}\label{Heff2}
 \begin{align}
 \langle\Pi_{1/2},J,p | H_\mathrm{sr} |\Pi_{1/2},J,p\rangle
 &=-\tfrac12 A + B(J+\tfrac12)^2
 +p(S_1+S_2)(2J+1)\,,
 \label{Heff2a}\\
 \langle\Pi_{3/2},J,p | H_\mathrm{sr} |\Pi_{3/2},J,p\rangle
 &=+\tfrac12 A + B(J+\tfrac12)^2 -2B\,,
 \label{Heff2b}\\
 \langle\Pi_{3/2},J,p | H_\mathrm{sr} |\Pi_{1/2},J,p\rangle
 &=\left[B +p S_2
 (J+\tfrac12)\right]
 \sqrt{(J-\tfrac12)(J+\tfrac32)}\,.
 \label{Heff2c}
 \end{align}
 \end{subequations}
Here in addition to the parameters $A$ and $B$ we have two parameters which
appear in the second order of perturbation theory via intermediate state(s)
$\Sigma_{1/2}$. The parameter $S_1$ corresponds to the cross term of the
perturbation theory in the spin-orbit and Coriolis interactions, while the
parameter $S_2$ is quadratic in the Coriolis interaction. Because of this
$S_1$ scales as $\alpha^2\mu$ and $S_2$ scales as $\mu^2$. It is easy to see
that the Hamiltonian $H_\mathrm{sr}$ describes limiting cases $|A|\gg B$ and
$|A|\ll B$ considered in Sec.\ \ref{analytic}.

The hyperfine part of the effective Hamiltonian is defined in the lowest order
of perturbation theory and has the form:
 \begin{subequations}\label{Heff3}
 \begin{align}
 \langle\Pi_{1/2},J,p | H_\mathrm{hf} |\Pi_{1/2},J,p\rangle
 &= C_F\left[2a-b-c + p(2J+1)d\right],
 \label{Heff3a}\\
 \langle\Pi_{3/2},J,p | H_\mathrm{hf} |\Pi_{3/2},J,p\rangle
 &=3C_F\left[2a+b+c\right],
 \label{Heff3b}\\
 \langle\Pi_{3/2},J,p | H_\mathrm{hf} |\Pi_{1/2},J,p\rangle
 &=-C_F\sqrt{(2J-1)(2J+3)}\,b\,,
 \label{Heff3c}
 \\
 C_F&\equiv [F(F+1)-J(J+1)-I(I+1)][8J(J+1)]^{-1}\,.
 \nonumber
 \end{align}
 \end{subequations}
Here we assume that only one nucleus has spin and include only magnetic dipole
hyperfine interaction.

The effective Hamiltonian described by Eqs.\ (\ref{Heff2},\ref{Heff3}) has 8
parameters. We use NIST values~\cite{66} for the fine structure
splitting $A$, rotational constant $B$, and magnetic hyperfine constants $a$,
$b$, $c$, $d$. Remaining two parameters $S_1$ and $S_2$ are found by
minimizing the \textit{rms} deviation between theoretical and experimental
$\Lambda$-doubling spectra.

In order to find sensitivity coefficients $Q_\alpha$ we calculate transition
frequencies for two values of $\alpha=\alpha_0\pm\delta$ near its physical
value $\alpha_0=1/137.035999679(94)$. The similar procedure is applied to
$Q_\mu$ at the physical value of the electron-to-proton mass ratio, $\mu_0 =
1/1836.15267247(80)$. We use scaling rules discussed above to recalculate
parameters of the effective Hamiltonian for different values of fundamental
constants. Then we use numerical differentiation to find respective
sensitivity coefficient.

%---------------------Figure 1
\begin{figure*}[t]
 \includegraphics[scale=0.80]{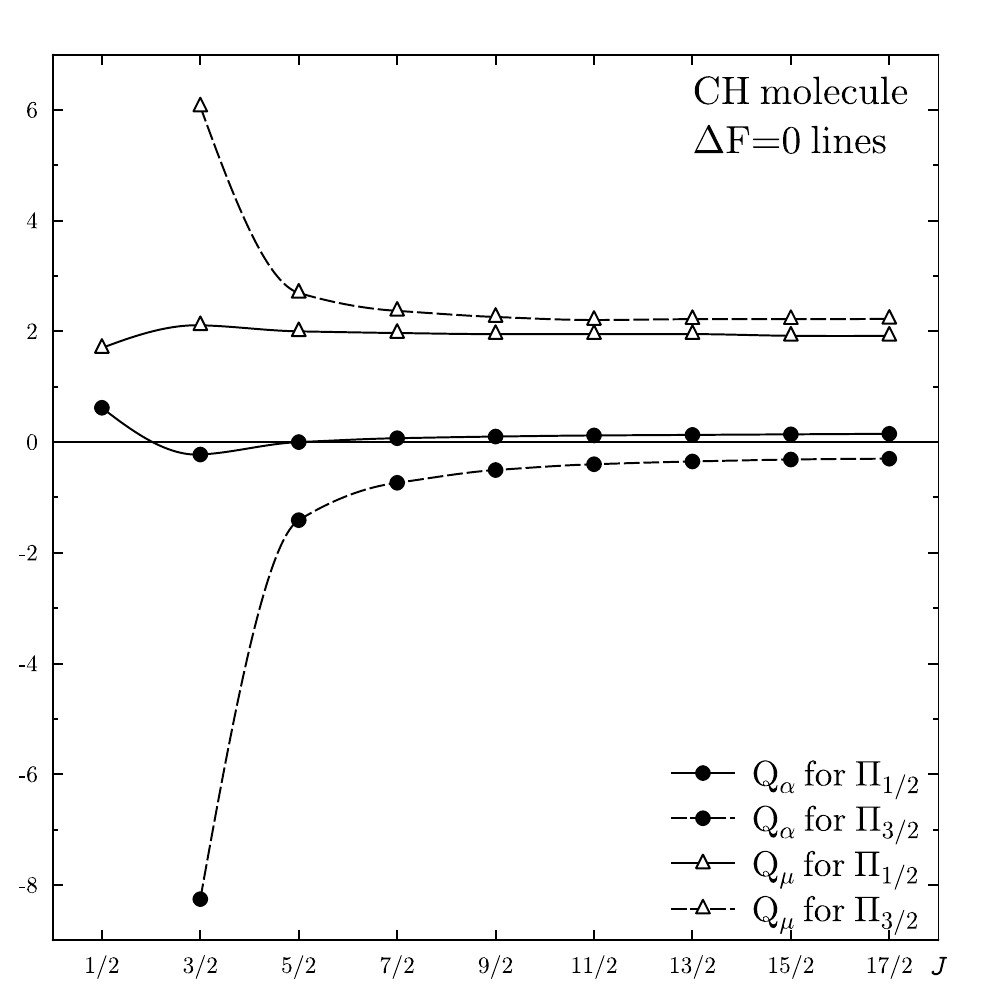}
 \hfill
 \includegraphics[scale=0.80]{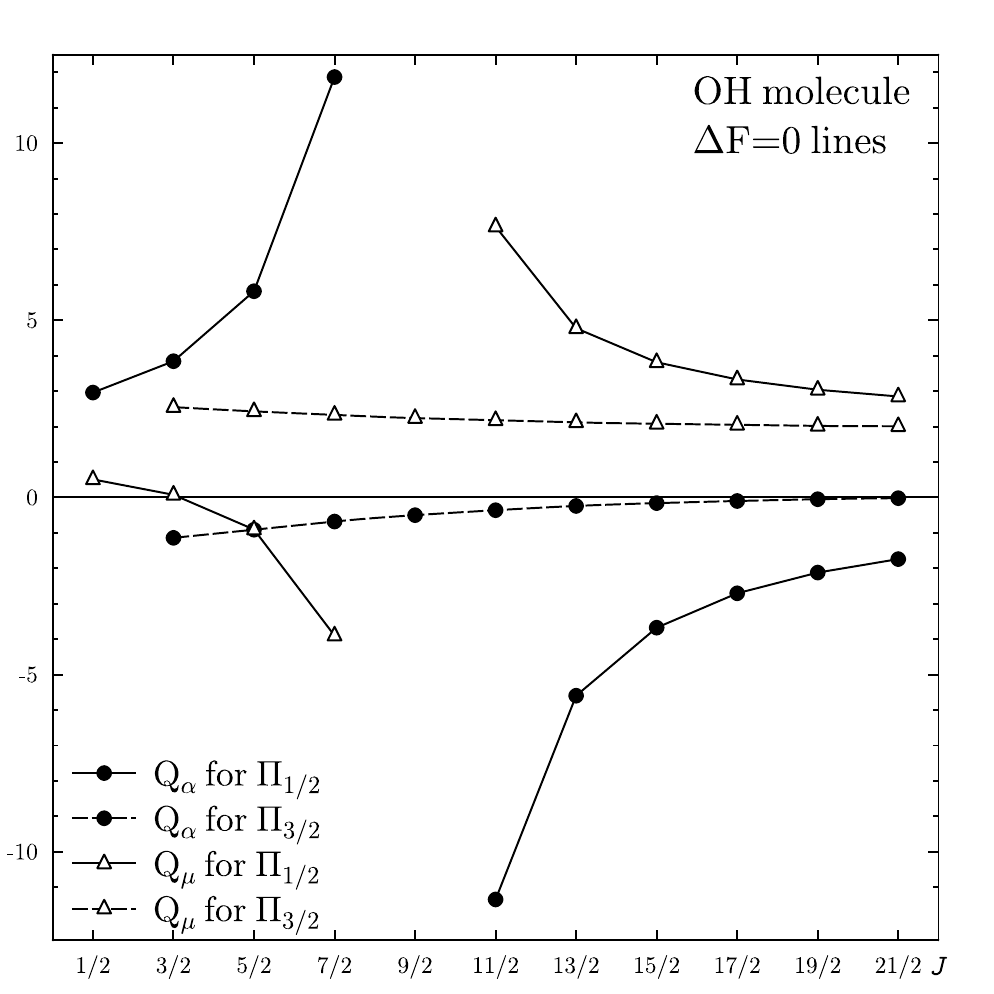}
 \caption{Sensitivity coefficients $Q_\alpha$ and $Q_\mu$ for
 $\Lambda$-doublet lines with
 $\Delta F=0$ in CH and OH. The difference between lines
 with $F=J+\tfrac12$ and $F=J-\tfrac12$ is too small to be seen.
 For the state $\Pi_{3/2}$ of OH the values for $J=\tfrac92$
 are too large to be shown on the plot. They are listed in \tref{tab-1}.}
 \label{fg1}
\end{figure*}

%----------------------------Figure 2
\begin{figure}[t]
 \includegraphics[scale=0.5]{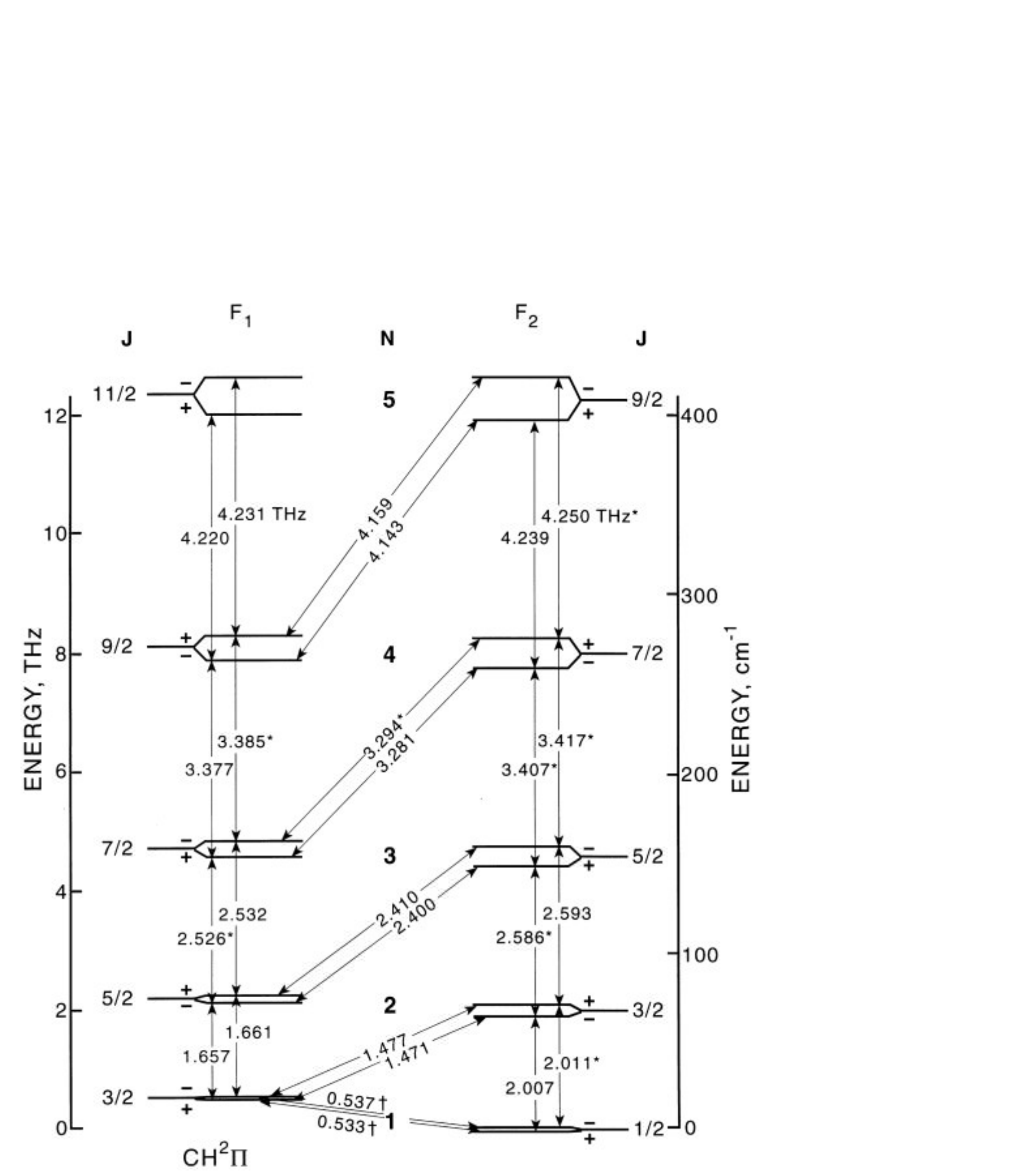}
 \caption{Rotational spectrum of CH from~\cite{21}. Vertical and
 diagonal arrows correspond to pure rotational and mixed transitions,
 respectively. $\Lambda$-doubling is not to scale.}
\label{fg2}
\end{figure}

\subsection{Sensitivity coefficients for $\Lambda$-doublet transitions in CH and OH}
\label{discussion}

In Ref.~\cite{51}, the method described in the previous section was applied to
$^{16}$OH, $^{12}$CH, $^{7}$Li$^{16}$O, $^{14}$N$^{16}$O, and
$^{15}$N$^{16}$O. The molecules CH and NO have ground state $^2\Pi_{1/2}$
($A>0$), while OH and LiO have ground state $^2\Pi_{3/2}$ ($A<0$). The ratio
$|A/B|$ changes from 2 for CH molecule~\cite{21}, to 7 for
OH~\cite{68}, and to almost a hundred for LiO and NO.
Therefore, LiO and NO definitely belong to the coupling case $a$. For OH
molecule we can expect transition from case $a$ for lower rotational states to
case $b$ for higher ones. Finally, for CH we expect intermediate coupling for
lower rotational states and coupling case $b$ for higher states.

Let us see how this scheme works in practice for the effective Hamiltonian
(\ref{Heff2},\ref{Heff3}). Fig.\ \ref{fg1} demonstrates $J$-dependence
of the sensitivity coefficients for CH and OH molecules. Both of them have
only one nuclear spin $I=\tfrac12$. For a given quantum number $J$, each
$\Lambda$-doublet transition has four hyperfine components: two strong
transitions with $\Delta F=0$ and $F=J\pm\tfrac12$ (for $J=\tfrac12$ there is
only one transition with $F=1$) and two weaker transitions with $\Delta F=\pm
1$. The hyperfine structure for OH and CH molecules is rather small and
sensitivity coefficients for all hyperfine components are very close. Because
of that Fig.\ \ref{fg1} presents only averaged values for strong
transitions with $\Delta F=0$.

We see that for large values of $J$ the sensitivity coefficients for both
molecules approach limit \eqref{doubling3} of the coupling case $b$. The
opposite limits (\ref{doubling1},\ref{doubling2}) are not reached for either
molecule even for smallest values of $J$. So, we conclude that the coupling
case $a$ is not realized. It is interesting that in Fig.\ \ref{fg1} the
curves for the lower states are smooth, while for upper states there are
singularities. For CH molecule this singularity takes place for the state
$\Pi_{3/2}$ near the lowest possible value $J=3/2$. A singularity for OH
molecule takes place for the state $\Pi_{1/2}$ near $J=9/2$.

These singularities appear because $\Lambda$-splitting turns to zero. As we
saw above, the sign of the splitting for the coupling case $a$ depends on the
sign of the constant $A$. The same sign determines which state $\Pi_{1/2}$, or
$\Pi_{3/2}$ lies higher. As a result, for the lower state the sign of the
splitting is the same for both limiting cases, but decoupling of the electron
spin $S$ for the upper state leads to the change of sign of the splitting. Of
course, these singularities are most interesting for our purposes, as they
lead to large sensitivity coefficients which strongly depend on the quantum
numbers. Note, that when the frequency of the transition is small, it becomes
sensitive to the hyperfine part of the Hamiltonian and the sensitivity
coefficients for hyperfine components may differ significantly. The
sensitivity coefficients of all hyperfine components of such $\Lambda$-lines
are given in \tref{tab-1}. We can see that near the singularities all
sensitivity coefficients are enhanced.

In addition to $\Lambda$-doublet transitions and purely rotational transitions
there are also mixed transitions between rotational states of $\Pi_{1/2}$ and
$\Pi_{3/2}$ states. The transition energy here includes the rotational and the
fine structure parts. Because of that, such transitions may have different
sensitivities to the variation of fundamental constants~\cite{22}.
As an example, \fref{fg2} shows mixed transitions in CH molecule. The
sensitivity coefficients are given in \tref{tab-2}. The isotopologue CD has
mixed transitions of lower frequencies and higher sensitivities~\cite{22}. 
Similar picture takes place for OH molecule.

%-------------------------------Table II
\begin{table}[t!]
\caption{Frequencies (GHz) and sensitivities of the rotational and mixed
transitions in CH.}
 \label{tab-2}
\begin{tabular}{ccdddd}
 \hline\hline
 \multicolumn{1}{c}{$N,J,p$}
 & \multicolumn{1}{c}{$N',J',p'$}
 & \multicolumn{1}{c}{$\nu_\mathrm{theor}$}
 & \multicolumn{1}{c}{$\nu_\mathrm{expt}$~\cite{77} }  
 & \multicolumn{1}{c}{$Q_\alpha$}
 & \multicolumn{1}{c}{$Q_\mu$}
 \\
 \hline\\[-3mm]
$ 1,\tfrac32,+ $&$ 1,\tfrac12,- $&  533.9 &  532.7   &  1.59 &  0.20 \\[1mm]
$ 1,\tfrac32,- $&$ 1,\tfrac12,+ $&  537.9 &  536.8   &  1.57 &  0.22 \\[1mm]
$ 2,\tfrac32,+ $&$ 1,\tfrac32,- $& 1477.2 & 1477.4   &  0.00 &  1.00 \\[1mm]
$ 2,\tfrac32,- $&$ 1,\tfrac32,+ $& 1470.6 & 1470.7   & -0.01 &  1.00 \\[1mm]
$ 2,\tfrac52,+ $&$ 1,\tfrac32,- $& 1663.0 & 1661.1   &  0.00 &  1.00 \\[1mm]
$ 2,\tfrac52,- $&$ 1,\tfrac32,+ $& 1658.8 & 1657.0   &  0.00 &  1.00 \\[1mm]
$ 2,\tfrac32,+ $&$ 1,\tfrac12,- $& 2011.8 & 2010.8   &  0.42 &  0.79 \\[1mm]
$ 2,\tfrac32,- $&$ 1,\tfrac12,+ $& 2007.8 & 2006.8   &  0.42 &  0.79 \\[1mm]
$ 2,\tfrac52,+ $&$ 2,\tfrac32,- $&  193.1 &  191.1   &  0.01 &  1.03 \\[1mm]
$ 2,\tfrac52,- $&$ 2,\tfrac32,+ $&  180.9 &  178.9   &  0.06 &  0.94 \\[1mm]
 \hline\hline
\end{tabular}
\end{table}

%---------------------- Figure 3
\begin{figure*}[t]
 \includegraphics[scale=1.0]{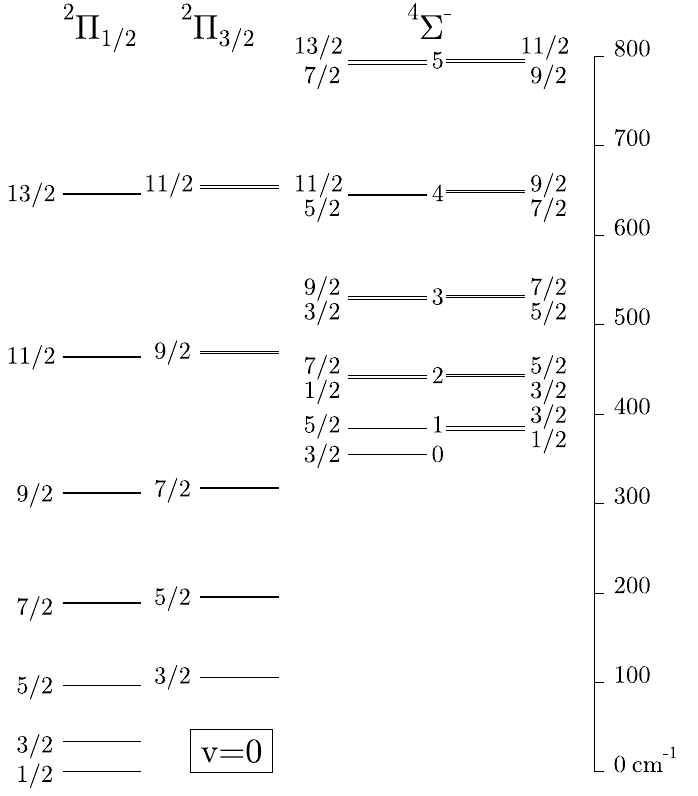}
 \qquad
 \includegraphics[scale=1.0]{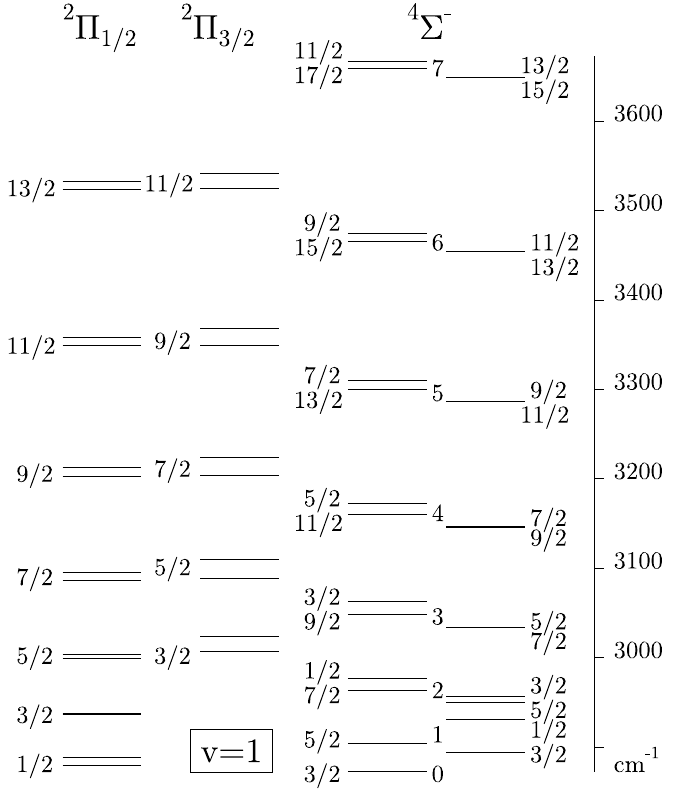}
 \caption{Spin-rotational levels of the three lowest electronic states of
 the molecule NH$^+$. Panels (a) and (b) correspond to vibrational states $v=0$
 and $v=1$ respectively. The energy levels are labeled with the quantum
 number $J$ for the $\Pi$ states and with $J$ and $N$ for the $\Sigma$ state.}
\label{fg3}
\end{figure*}

The molecule NH$^+$ is isoelectronic to CH and also has ground state
$^2\Pi_{1/2}$. However, there is an important difference: for NH$^+$ the first
excited state $^4\Sigma^-$ lies only 340 cm$^{-1}$ above the ground state
\cite{47,40}. The spin-orbit interaction
between these states leads to strong perturbations of the rotational structure
and of the $\Lambda$-doublet splittings and to an additional enhancement of
the sensitivity coefficients~\cite{5}. The spectrum of NH$^+$ is
shown in Fig.\ \ref{fg3}. The effective Hamiltonian is similar to the one
considered above with two additional terms describing interaction between the
$^2\Pi$ and $^4\Sigma$ states~\cite{47}: 
 \begin{subequations}\label{Heff4}
 \begin{align}
 \langle^2\Pi_{3/2},J,p| H_\mathrm{so} |^4\Sigma_{3/2}^-,J,p\rangle
 &=-\frac12 \zeta_{3/2}\,,
 \label{Heff4a}\\
 \langle^2\Pi_{1/2},J,p| H_\mathrm{so} |^4\Sigma_{1/2}^-,J,p\rangle
 &=-\frac{1}{2\sqrt{3}} \zeta_{1/2}\,.
 \label{Heff4b}
 \end{align}
 \end{subequations}
Obviously, the parameters $\zeta_{1/2}$ and $\zeta_{3/2}$ scale as $\alpha^2$.
As mentioned above, for the NH$^+$ molecule the splitting between $\Sigma$ and
$\Pi$ states $\Delta E_{\Sigma\Pi}$ is only about 340 cm$^{-1}$. This
splitting includes three contributions: the non-relativistic electronic energy
difference, the relativistic corrections ($\sim\alpha^2 Z^2$) and the
difference in the zero point vibrational energies for the two states
($\sim\mu^{1/2}$). Note that the accidental degeneracy of these levels for
NH$^+$ means that the first contribution is anomalously small. Because of
that, the other two contributions can not be neglected and modify the scaling
of $\Delta E_{\Sigma\Pi}$ with fundamental constants. This effect has to be
taken into account in the calculations of the sensitivity coefficients~\cite{5}.

\section{Linear polyatomic radicals in the $\Pi$ ground state: C$_3$H}
\label{sect-4}

The linear form of the molecule C$_3$H ($l$-C$_3$H) is similar to the molecule
NH$^+$: it also has the ground state $^2\Pi_{1/2}$ and two closely lying
states $^2\Pi_{3/2}$ and $^2\Sigma_{1/2}^+$. Here the quasi degeneracy of the
$\Pi$ and $\Sigma$ states is not accidental, but is caused by the
Renner-Teller interaction. In the following section we briefly recall the
theory of the Renner-Teller effect in polyatomic linear molecules~\cite{89,39}.

%------------------------------------Table III
\begin{table}[t!]
  \caption{Low lying energy levels for the bending mode $\omega_v=589$
  cm$^{-1}$ of $l$-C$_3$H molecule and their sensitivities $q_\alpha$
  and $q_\mu$ to the variation of $\alpha$ and $\mu$ respectively. 
  $\Delta$ is the distance from the ground state.
  All values are in cm$^{-1}$.}
 \label{tab-3}
 \begin{tabular}{clcddcddcd}
 \hline\hline
 \multicolumn{1}{c}{$v_\mathrm{nom}$}
 &\multicolumn{1}{c}{$\langle v\rangle$}
 &\multicolumn{1}{c}{$K$}
 &\multicolumn{1}{c}{$\Omega$}
 &\multicolumn{1}{c}{$\langle\Lambda\rangle$}
 &\multicolumn{1}{c}{$E$}
 &\multicolumn{2}{c}{$\Delta$}
 &\multicolumn{1}{c}{$q_\mu$}
 &\multicolumn{1}{c}{$q_\alpha$}
 \\
&&&&&&\multicolumn{1}{c}{\cite{49}}
 &\multicolumn{1}{c}{\cite{81}}
 \\
 \hline
  0 & 1.22  & 1 & 0.5 & 0.50 & 367.9 &  0.0   &  0.0 & 187.8  &-14.6  \\
  0 & 1.35  & 1 & 1.5 & 0.46 & 381.9 & 13.9   & 14.0 & 187.8  & 13.3  \\
  1 & 2.32  & 0 & 0.5 &-0.01 & 394.2 & 26.3   & 27.0 & 197.3  & -0.4  \\
  1 & 3.57  & 2 & 1.5 & 0.21 & 597.7 &229.7   &226.0 & 300.3  & -6.1  \\
  1 & 3.65  & 2 & 2.5 & 0.19 & 603.5 &235.5   &232.0 & 300.3  &  5.5  \\
 \hline\hline
\end{tabular}
\end{table}

\subsection{Renner-Teller effect}
\label{sec_basics}

The total molecular angular momentum of the polyatomic molecule $\bm{J}$
includes the vibrational angular momentum $\bm{G}$ associated with the twofold
degenerate bending vibration mode(s):
$\bm{J}=\bm{N}+\bm{S}=\bm{R}+\bm{G}+\bm{L}+\bm{S}$, where $\bm{R}$ describes
rotation of the molecule as a whole and is perpendicular to the molecular axis
$\zeta$. Other momenta have nonzero $\zeta$-projections: $\langle
G_\zeta\rangle =l$,  $\langle L_\zeta\rangle =\Lambda$, $\langle
N_\zeta\rangle =K=l+\Lambda$, and $\langle J_\zeta\rangle=\Omega$.

Suppose we have $\Pi$ electronic state $|\Lambda=\pm 1\rangle$ and $v=1$
vibrational state of a bending mode $|l=\pm 1\rangle$. All together there are
4 states $|\Lambda=\pm 1\rangle |l=\pm 1\rangle$. We can rewrite them as one
doublet $\Delta$ state $|K=\pm 2\rangle$ and states $\Sigma^+$ and $\Sigma^-$.
In the adiabatic approximation all four states are degenerate. Renner~\cite{89} 
showed that the states with the same quantum number $K=l+\Lambda$ strongly
interact, so the $\Sigma^+$ and $\Sigma^-$ states repel each other, while
the $\Delta$ doublet in the first approximation remains unperturbed. We are
particularly interested in the case when one of the $\Sigma$ levels is pushed
close to the ground state $v=0$. This is what takes place in the $l$-C$_3$H
molecule~\cite{104,45,14}. 

Consider a linear polyatomic molecule with the unpaired electron in the
$\pi_\xi$ state in the molecular frame $\xi,\eta,\zeta$. Obviously, the
bending energy is different for bendings in $\xi\zeta$ and in $\eta\zeta$
planes: $V_\pm=\tfrac12 k_\pm\chi^2$ (here $\chi$ is the supplement to the
bond angle). That means that the electronic energy depends on the angle $\phi$
between the electron and nuclear planes:
 \begin{align}\label{Hprim}
 H'=V'\cos 2\phi\,,
 \end{align}
where $2V'=V_+-V_-=k'\chi^2$. There is no reason for $V'$ to be small, so
$k'\sim k_\pm\sim 1$ a.u.\ and to a first approximation $k'$ does not depend
on $\alpha$ and $\mu$.

As long as interaction \eqref{Hprim} depends on the relative angle between the
electron and the vibrational planes, it changes the angular quantum numbers as
follows: $\Delta\Lambda=-\Delta l=\pm2$ and $\Delta K=0$. This is exactly what
is required to produce splitting between the $\Sigma^+$ and $\Sigma^-$ states
with $v=1$ as discussed above.

%----------------------------Table IV
\begin{table}[tbh]
\caption{$l$-C$_3$H
sensitivity coefficients for the transitions between states from \tref{tab-3}
  and for parameters $A_\mathrm{eff}$ and $\Delta E_{\Sigma\Pi}$ defined
  by \eqref{Aeff} and \eqref{DelE} respectively. Frequencies are in cm$^{-1}$.}
\label{tab-4}
 \begin{tabular}{cdcddddddd}
 \hline\hline
 &&&&\multicolumn{3}{c}{Fit to  \cite{81}}
 &\multicolumn{3}{c}{\, Fit to  \cite{14}} \\
 \multicolumn{1}{c}{$K$}
 &\multicolumn{1}{c}{$\Omega$}
 &\multicolumn{1}{c}{$K'$}
 &\multicolumn{1}{c}{$\Omega'$}
 &\multicolumn{1}{c}{$\omega$}
 &\multicolumn{1}{c}{$Q_\mu$}
 &\multicolumn{1}{c}{$Q_\alpha$}
 &\multicolumn{1}{c}{$\omega$}
 &\multicolumn{1}{c}{$Q_\mu$}
 &\multicolumn{1}{c}{$Q_\alpha$}
 \\
 \hline
  1 & 0.5  & 1 & 1.5 &  13.9 & 0.00&  2.00 &  14.4 & 0.00&  2.00  \\
  1 & 1.5  & 0 & 0.5 &  12.4 & 0.78& -1.11 &  13.3 & 0.77& -1.07  \\
  0 & 0.5  & 2 & 1.5 & 203.5 & 0.51& -0.03 & 204.4 & 0.51& -0.03  \\
  2 & 1.5  & 2 & 2.5 &   5.8 & 0.00&  2.00 &   6.0 & 0.00&  2.00  \\
 \multicolumn{4}{c}{$A_\mathrm{eff}$}
                     &  13.9 & 0.00&  2.00 &  14.4 & 0.00&  2.00  \\
 \multicolumn{4}{c}{$\Delta E_{\Sigma\Pi}$}
                     &  19.4 & 0.50&  0.00 &  20.5 & 0.50&  0.00  \\

 \hline\hline
\end{tabular}
\end{table}

Interaction \eqref{Hprim} also mixes different vibrational levels with $\Delta
v=\pm2,\pm4,\dots$. Thus, we have, for example, the nonzero ME $\langle
0,0,1,1|H'|2,2,-1,1\rangle$ between states $|v,l,\Lambda,K\rangle$. Such
mixings reduce effective value of the quantum number $\Lambda$ and, therefore,
reduce the spin-orbital splitting between the $\Pi_{1/2}$ and $\Pi_{3/2}$
states~\cite{81},
 \begin{align}\label{Aeff}
 H_\mathrm{so} &\equiv A_\mathrm{eff}\Lambda\Sigma\,,
 \quad A_\mathrm{eff}=A\Lambda_\mathrm{eff}/\Lambda\,.
 \end{align}

Let us define the model more accurately. Following~\cite{81} we
write the Hamiltonian as:
 \begin{align}\label{Ham1}
 H &= H_e + T_v + A L_\zeta S_\zeta\,.
 \end{align}
Here the ``electronic'' part $H_e$ includes all degrees of freedom except for
the bending vibrational mode and spin. For $l$-C$_3$H there are two bending
modes, but for simplicity we include the second bending mode in $H_e$ too.
Electronic MEs in the $|\Lambda\rangle$ basis have the form:
 \begin{subequations}\label{Ham2}
 \begin{align}
 \label{Ham2a}
 \langle \pm1|H_e|\pm1 \rangle &= \frac{V_+ + V_-}{2} = \frac{k}{2}\chi^2\,,
 \\ \label{Ham2b}
 \langle \pm1|H_e|\mp1 \rangle &= \frac{k'}{2}\chi^2 \exp{(\mp 2i\phi)} \,.
 \end{align}
 \end{subequations}
Here $\chi$ and $\phi$ are the vibrational coordinates for the bending mode.
Kinetic energy in these coordinates has the form:
 \begin{align}
 \label{Kin}
 T_v &= -\frac{1}{2MR^2} \left(\frac{\partial^2}{\partial\chi^2}
 +\frac{1}{\chi}\frac{\partial}{\partial\chi}
 +\frac{1}{\chi^2}\frac{\partial^2}{\partial\phi^2}
 \right)\,.
 \end{align}

We can use the basis set of 2D harmonic functions in polar coordinates
$\rho=\chi R$ and $\phi$ for the mass $M$ and the force constant $k$:
 \begin{align}
 \label{Oscil1}
 \psi_{v,l}(\rho,\phi) &= R_{v,l}(\rho)\frac{1}{\sqrt{2\pi}}\exp{(il\phi)}\,.
 \end{align}
It is important that the radial functions are orthogonal only for the same
$l$:
 \begin{align}
 \label{Oscil2}
 \langle R_{v',l}|R_{v,l}\rangle = \delta_{v',v}\,.
 \end{align}
This allows for the nonzero MEs between states with different quantum number
$l$. By averaging operator \eqref{Ham1} over vibrational functions we get:
 \begin{multline}\label{Ham4}
 \langle v',l'|H_e+T_v|v,l\rangle
 = \bigl[\omega_v(v+1)+A\Lambda S_\zeta \bigr] \delta_{v',v} \delta_{l',l}
 \\
 +\frac12 \langle R_{v'l'}|k'\chi^2|R_{vl}\rangle
 \exp{(\mp 2i\phi)} \delta_{l',l\pm2} \,.
 \end{multline}
The exponent here ensures the selection rule $\Lambda'=\Lambda\mp2$ for the
quantum number $\Lambda$ when we calculate MEs for the rotating molecule.

%----------------------------- Figure 4
\begin{figure*}[t]
 \vspace*{-20mm}
 \includegraphics[scale=0.6]{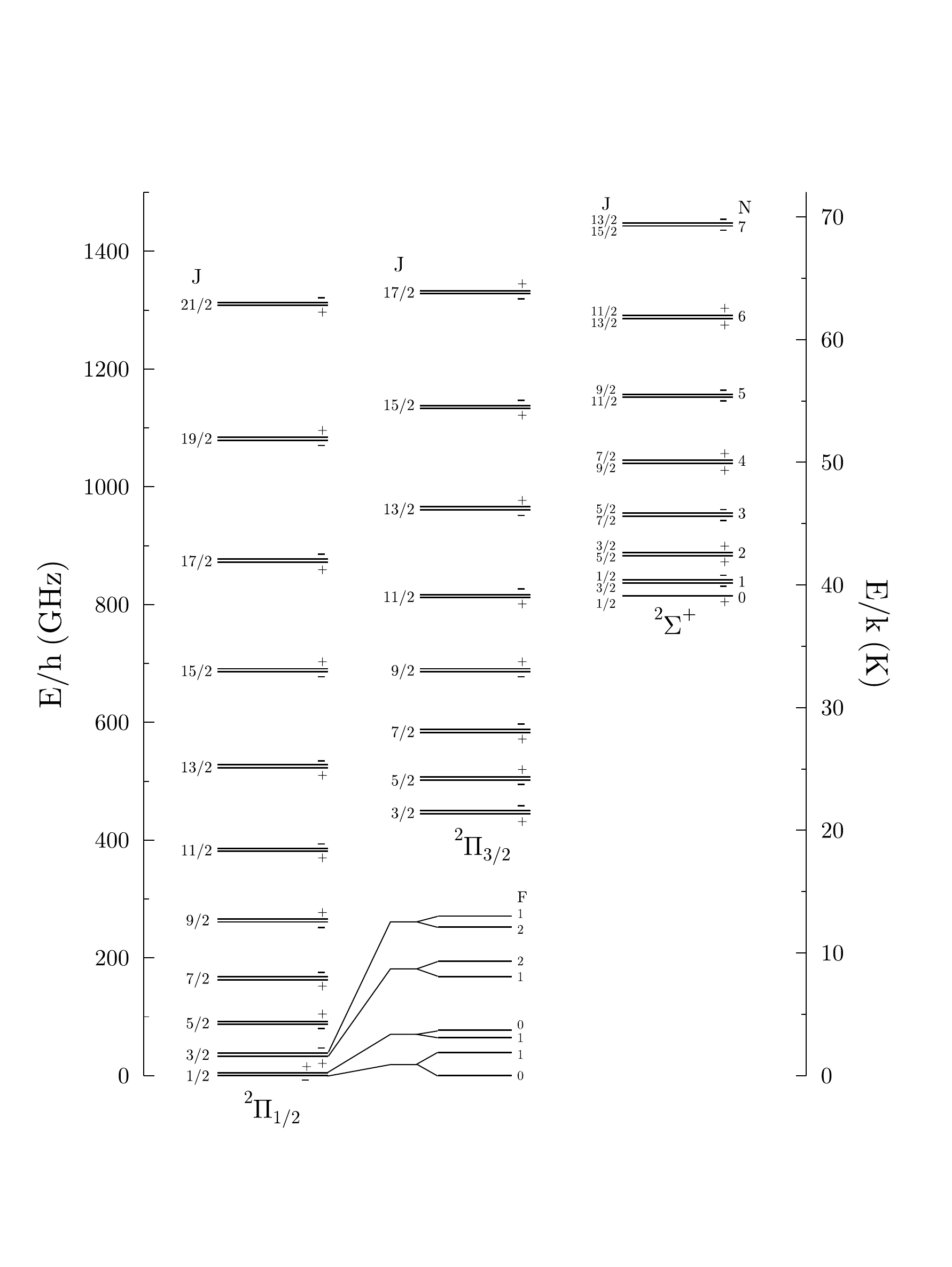}
 \vspace*{-20mm}
 \caption{Spin-rotational levels of the three lowest vibronic states of
 the molecule $l$-C$_3$H. $K$-doubling is indicated schematically,
hyperfine structure is shown only for the two lowest $K$-doublets. Due to
a strong Renner-Teller effect the component $^2\Sigma^+$ of the excited
bending state $\nu_4$(CCH bending) is shifted towards lower energies, $\sim
29$ cm$^{-1}$ above the zero-level of the ground state $^2\Pi_{1/2}$. }
\label{fg4}
\end{figure*}

\subsection{Molecule $l$-C$_3$H}
\label{sec-lC3H}

We solve the eigenvalue problem for Hamiltonian \eqref{Ham1} using the basis
set of the 2D-harmonic oscillator. Our model Hamiltonian has only 3
parameters, namely $\omega_v$, $A$, and the dimensionless Renner-Teller
parameter $\cal E$: $k'={\cal E}k$. The values for $\omega_v$ and $A$ for
$l$-C$_3$H are given in \cite{81}. We varied the Renner-Teller
parameter $\cal E$ to fit five lowest levels for the given bending mode:
$\Pi_{1/2}$, $\Pi_{3/2}$, $\Sigma_{1/2}$, $\Delta_{3/2}$, and $\Delta_{5/2}$.
The optimal value appeared to be ${\cal E}=0.788$. The results are presented
in \tref{tab-3}. The first two columns give nominal vibrational quantum number
$v$ and its actual average value. We see that the Renner-Teller term in
\eqref{Ham4} strongly mixes vibrational states. This mixing also affects
$\langle\Lambda\rangle$ and decreases spin-orbital splittings as explained by
\Eref{Aeff}.

%--------------------------Table V
\begin{table}[t!]
\caption{Frequencies (MHz), sensitivity coefficients, and reduced MEs  (a.u.)
for some $K$-doubling transitions in $\Pi_{1/2}$ state of the $l$-C$_3$H molecule. }
\label{tab-5}
\begin{tabular}{crddd}
\hline\hline
 \multicolumn{1}{c}{$J\,F'\!p'\!,Fp$}
&\multicolumn{1}{c}{$\omega$} &\multicolumn{1}{c}{$Q_\alpha$}
&\multicolumn{1}{c}{$Q_\mu$}
&\multicolumn{1}{c}{$||D||^2$}\\
\hline\\[-3mm]
$\frac{1}{2}\,{1+,\,\,0-} $&   52.37 &  0.66(2) & 1.7(2)   & 0.333 \\[1pt]
$\frac{1}{2}\,{0+,\,\,1-} $&   39.12 &  0.20(2) & 1.9(2)   & 0.333 \\[1pt]
$\frac{1}{2}\,{1+,\,\,1-} $&   34.93 & -0.02(2) & 2.0(2)   & 0.667 \\[3pt]
$\frac{3}{2}\,{1-,\,\,1+} $&   85.55 &  0.65(2) & 1.7(1)   & 0.166 \\[1pt]
$\frac{3}{2}\,{2-,\,\,1+} $&   78.60 &  0.55(2) & 1.7(1)   & 0.033 \\[1pt]
$\frac{3}{2}\,{1-,\,\,2+} $&   75.23 &  0.43(2) & 1.8(1)   & 0.033 \\[1pt]
$\frac{3}{2}\,{2-,\,\,2+} $&   68.29 &  0.30(2) & 1.8(1)   & 0.299 \\[3pt]
$\frac{5}{2}\,{2+,\,\,2-} $&  107.19 &  0.95(2) & 1.5(1)   & 0.132 \\[1pt]
$\frac{5}{2}\,{3+,\,\,2-} $&   98.97 &  0.89(2) & 1.5(1)   & 0.009 \\[1pt]
$\frac{5}{2}\,{2+,\,\,3-} $&   98.83 &  0.82(2) & 1.6(1)   & 0.009 \\[1pt]
$\frac{5}{2}\,{3+,\,\,3-} $&   90.61 &  0.75(2) & 1.6(1)   & 0.188 \\[3pt]
$\frac{7}{2}\,{3-,\,\,3+} $&  112.38 &  1.63(2) & 1.2(1)   & 0.105 \\[1pt]
$\frac{7}{2}\,{4-,\,\,4+} $&   96.07 &  1.56(2) & 1.2(1)   & 0.136 \\[3pt]
$\frac{9}{2}\,{4+,\,\,4-} $&   95.75 &  3.22(4) & 0.36(7)  & 0.086 \\[1pt]
$\frac{9}{2}\,{5+,\,\,5-} $&   79.63 &  3.45(4) & 0.23(7)  & 0.105 \\[3pt]
$\frac{11}{2}\,{5-,\,\,5+}$&   52.81 &  9.1 (6) &-2.6 (3)  & 0.072 \\[1pt]
$\frac{11}{2}\,{6-,\,\,6+}$&   36.85 & 12.1 (6) &-4.1 (3)  & 0.085 \\[3pt]
$\frac{13}{2}\,{6-,\,\,6+}$&   20.25 &-34.  (2) &19.  (2)  & 0.062 \\[1pt]
$\frac{13}{2}\,{7-,\,\,7+}$&   36.06 &-18.  (2) &11.  (2)  & 0.071 \\[3pt]
$\frac{15}{2}\,{7+,\,\,7-}$&  126.59 & -7.6 (2) & 5.8 (4)  & 0.054 \\[1pt]
$\frac{15}{2}\,{8+,\,\,8-}$&  142.24 & -6.5 (2) & 5.3 (4)  & 0.061 \\[1pt]
$\frac{17}{2}\,{8-,\,\,8+}$&  268.76 & -4.7 (1) & 4.4(3)   & 0.047 \\[1pt]
$\frac{17}{2}\,{9-,\,\,9+}$&  284.25 & -4.3 (1) & 4.2(3)   & 0.053 \\[1pt]
$\frac{19}{2}\,{9+,\,\,9-}$&  448.75 & -3.59(7) & 3.8(3)   & 0.042 \\[1pt]
$\frac{19}{2}\,{10+,\,\,10-}$&464.07 & -3.39(7) & 3.7(3)   & 0.046 \\[1pt]
$\frac{21}{2}\,{10-,\,\,10+}$&668.02 & -2.97(6) & 3.5 (3)  & 0.038 \\[1pt]
$\frac{21}{2}\,{11-,\,\,11+}$&683.18 & -2.85(6) & 3.4 (3)  & 0.041 \\[1pt]
\hline
\end{tabular}
\end{table}

The last two columns in \tref{tab-3} give dimensional sensitivity coefficients
$q_\mu$ and $q_\alpha$ in cm$^{-1}$:
$$ \Delta E = q_\alpha \frac{\Delta \alpha}{\alpha}
  + q_\mu \frac{\Delta \mu}{\mu}\,.$$
To estimate them we assumed that the parameters scale in a following way:
$\omega_v\sim \mu^{1/2}$, $A\sim\alpha^2$, and $\cal E$ does not depend on
$\alpha$ and $\mu$. The dimensionless sensitivity coefficients
\eqref{Q-factors} for the transitions $\omega_{i,k}=E_k-E_i$ can be found as:
 $$Q_{i,k}=(q_k-q_i)/\omega_{i,k}\,.$$
In \tref{tab-4} these coefficients are calculated for the same set of
parameters as in \tref{tab-3} and for the slightly different parameters which
better fit experimental frequencies from~\cite{14}. We see that the
sensitivity coefficients are practically the same for both sets.

For the two fine structure transitions, $\Pi_{1/2}\rtw \Pi_{3/2}$ and
$\Delta_{3/2}\rtw \Delta_{5/2}$, we get sensitivities $Q_\mu=0$ and
$Q_\alpha=2$. This may seem strange as the fine structure is significantly
reduced by the Renner-Teller mixing: the fine-structure parameter is 29
cm$^{-1}$ and the splitting between $\Pi_{1/2}$ and $\Pi_{3/2}$ is only 13.9
cm$^{-1}$. According to \eqref{Aeff} the mixing reduces the splitting.
However, this effect depends on the dimensionless Renner-Teller parameter
$\cal E$ and does not depend on $\mu$ and $\alpha$. Consequently, \textit{the
effective parameter $A_\mathrm{eff}$ depends on fundamental constants in the
same way as initial parameter $A$.}

For the high frequency transition $\Sigma_{1/2}\rtw \Delta_{3/2}$, where the
spin-orbital energy can be neglected, we get $Q_\mu=0.5$ and $Q_\alpha=0$.
These results are expected, because our model has only two dimensional
parameters: vibrational frequency, which is proportional to $\mu^{1/2}$ and
the fine structure parameter $A$, which scales as $\alpha^2$. Even though our
vibrational spectrum is far from that of a simple harmonic oscillator, the
non-diagonal MEs \eqref{Ham4} of the Hamiltonian \eqref{Ham1} still scale as
$\mu^{1/2}$. Therefore, if we neglect spin-orbital splittings, we get
$Q_\mu=1/2$ for all transitions. The only transition in \tref{tab-4} where the
spin-orbital energy and vibrational energy are close to each other is the
$\Pi_{3/2}\rtw \Sigma_{1/2}$ transition. The resultant frequency is roughly
half of the vibrational energy difference between the $\Pi$ and $\Sigma$
states. This leads to $Q_\mu\approx 1$ and $Q_\alpha\approx -1$.

%-------------------------------------Table VI
\begin{table}[t!]
\caption{Frequencies (MHz), sensitivity coefficients, and reduced MEs  (a.u.)
for some $K$-doubling transitions in $\Pi_{3/2}$ state of the $l$-C$_3$H molecule.}
\label{tab-6}
\begin{tabular}{crddd}
\hline\hline
 \multicolumn{1}{c}{$J\,F'\!p'\!,Fp$}
 &\multicolumn{1}{c}{$\omega$}
 &\multicolumn{1}{c}{$Q_\alpha$} &\multicolumn{1}{c}{$Q_\mu$}
 &\multicolumn{1}{c}{$||D||^2$}\\
\hline\\[-3mm]
$\frac{3}{2}\,{1-,1+} $&     5.61 & -2.63(8) &  3.2 (2) & 1.493 \\[1pt]
$\frac{3}{2}\,{2-,1+} $&    18.50 &  0.49(8) &  1.7 (2) & 0.299 \\[1pt]
$\frac{3}{2}\,{1-,2+} $&    -7.30 &  5.28(8) & -0.6 (2) & 0.299 \\[1pt]
$\frac{3}{2}\,{2-,2+} $&     5.58 & -2.63(8) &  3.2 (2) & 2.688 \\[3pt]
$\frac{5}{2}\,{2+,2-} $&    22.24 & -2.60(8) &  3.2 (2) & 1.186 \\[1pt]
$\frac{5}{2}\,{3+,2-} $&    31.50 & -1.35(8) &  2.6 (2) & 0.085 \\[1pt]
$\frac{5}{2}\,{2+,3-} $&    12.88 & -5.67(8) &  4.6 (2) & 0.085 \\[1pt]
$\frac{5}{2}\,{3+,3-} $&    22.15 & -2.60(8) &  3.2 (2) & 1.694 \\[3pt]
$\frac{7}{2}\,{3-,3+} $&    54.92 & -2.57(8) &  3.2 (2) & 0.943 \\[1pt]
$\frac{7}{2}\,{4-,4+} $&    54.76 & -2.57(8) &  3.2 (2) & 1.223 \\[5pt]
$\frac{9}{2}\,{+-} $&  108.13 & -2.50(8) &  3.1 (2) & 1.230 \\[1pt]
$\frac{11}{2}\,{-+}$&  185.99 & -2.46(8) &  3.1 (2) & 1.007 \\[1pt]
$\frac{39}{2}\,{-+}$& 4266.17 & -2.9 (1) &  2.53(8) & 0.224 \\[1pt]
$\frac{41}{2}\,{+-}$& 4553.04 & -3.5 (1) &  2.42(5) & 0.208 \\[1pt]
$\frac{43}{2}\,{-+}$& 4663.43 & -4.6 (2) &  2.2 (1) & 0.192 \\[1pt]
$\frac{45}{2}\,{+-}$& 4377.16 & -7.5 (2) &  1.4 (3) & 0.174 \\[1pt]
$\frac{47}{2}\,{-+}$& 3097.96 &-19.0 (4) & -2.3 (9) & 0.149 \\[1pt]
$\frac{49}{2}\,{-+}$&  909.06 &132.  (2) & 53.(8)   & 0.103 \\[1pt]
$\frac{51}{2}\,{-+}$&19813.69 & -3.11(5) & -1.6 (4) & 0.116 \\[1pt]
\hline
\end{tabular}
\end{table}

The spectrum of the $l$-C$_3$H molecule is shown on \fref{fg4}. The
effective Hamiltonian for the rotating molecule is similar to that of the
NH$^+$ molecule. It includes the effective fine-structure parameter
$A_\mathrm{eff}$ and the energy difference between the $\Sigma$ and $\Pi$
states,
 \begin{equation}\label{DelE}
 \Delta E_{\Sigma\Pi}=E(\Sigma^+)-\frac{E(\Pi_{1/2})+E(\Pi_{3/2})}{2}\,.
 \end{equation}
Numerical values for these parameter are obtained from the fit to the
experimental transition frequencies. Here we only need to determine the
dependence of these parameters on fundamental constants. \tref{tab-4} shows
that $A_\mathrm{eff}\sim \alpha^2$ and $\Delta E_{\Sigma\Pi}\sim \mu^{1/2}$.
Once again, this is because the Renner-Teller mixing depends on the
dimensionless parameter $\cal E$ and \textit{does not} depend on $\alpha$ and
$\mu$. Calculated sensitivity coefficients for the $K$-doublet transitions of
the $l$-C$_3$H molecule are listed in Tables \ref{tab-5} and \ref{tab-6}. The
results for the mixed transitions can be found in \cite{49}.

\section{Tunneling modes in polyatomic molecules}
\label{sect-5}

%------------------------- Figure 5
\begin{figure*}[t]
 \includegraphics[scale=0.80]{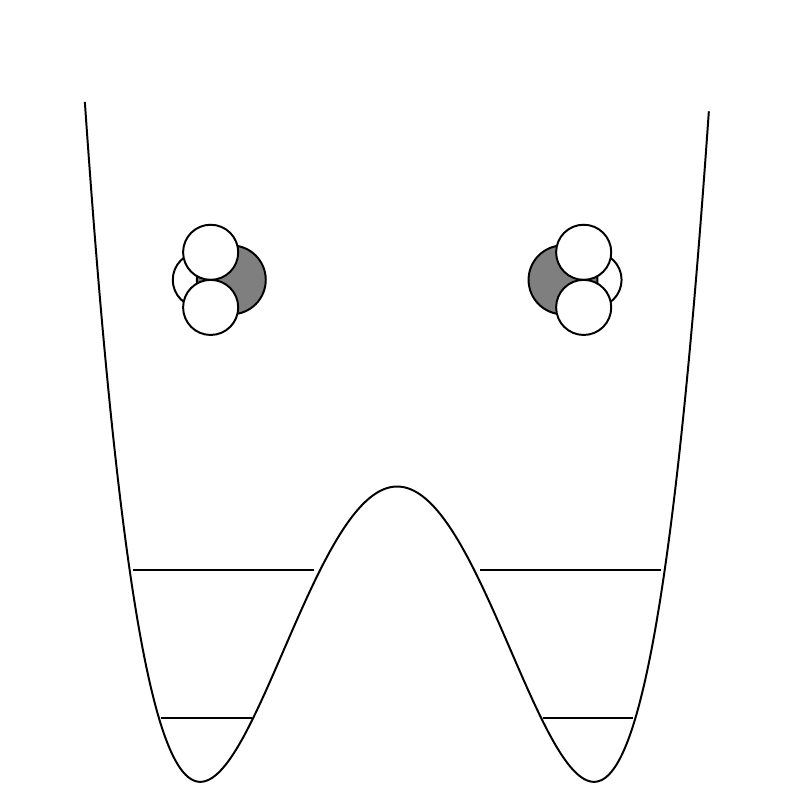}
 \caption{Potential for the tunneling (umbrella) mode of the NH$_3$ molecule. Two
 lowest vibrational levels lie below the barrier.}
\label{fg5}
\end{figure*}

In this section we consider non linear and non planar polyatomic molecules.
Such molecules generally have more than one equivalent potential minimum. If
the barriers between these minima are not too high the molecule can tunnel
between them. Ammonia (NH$_3$) is the best known textbook example of a
nonrigid molecule (see \fref{fg5}). Interestingly, this molecule is also
one of the most abundoned polyatomic molecules in the interstellar medium.
Other important for astrophysics molecules with tunneling include hydronium
(H$_3$O$^+$), peroxide (H$_2$O$_2$), methanol (CH$_3$OH), and methylamine
(CH$_3$NH$_2$). We will briefly discuss all of them below. All these molecules
include only light atoms with $Z\le 8$ and have singlet electronic ground
states. Thus we can neglect relativistic corrections and assume that all
discussed transitions have $Q_\alpha=0$.

It is clear that tunneling frequencies should strongly depend on the nuclear
masses, and we can expect large sensitivity coefficients
$Q_{\mu,\mathrm{tun}}$. They can be found using the semi-classical
Wentzel-Kramers-Brillouin (WKB) approximation. Following~\cite{111} 
we can write the ground state tunneling frequency in atomic units
($\hbar=|e|=m_e=1$) as:
 \begin{equation} \label{Eq:w_tun}
 \omega_{\rm tun} \approx \frac{2 E_0}{\pi}\,e^{-S},
 \end{equation}
where $S$ is the action over classically forbidden region and $E_0$ is the
ground state vibrational energy calculated from the bottom of the well $U_{\rm
min}$. If the barrier is high enough the harmonic approximation gives $2
E_0=\omega_v$, where $\omega_v$ is the observed vibrational frequency. In this
case \Eref{Eq:w_tun} allows to find action $S$ from experimentally known
frequencies $\omega_{\rm tun}$ and $\omega_v$. For lower barriers we need to
know the shape of the potential to estimate $E_0$. The examples of these two
limiting cases are ammonia and hydronium, where tunneling frequencies are 0.8
cm$^{-1}$ and 55 cm$^{-1}$ respectively.

The action $S$ depends on the tunneling mass, which in atomic units is
proportional to $\mu^{-1}$. Differentiating \eqref{Eq:w_tun} over $\mu$ we get~\cite{32,53}
\begin{equation}
 Q_{\mu,\mathrm{tun}} \approx \frac{1+S}{2} + \frac{S\, E_0}{2(\Delta
 U-E_0)}\,,
 \label{Eq:Q_WKB}
\end{equation}
where $\Delta U=U_{\rm max}-U_{\rm min}$ is the barrier hight. Numerical
solution of the Schr\"odinger equation for realistic potentials agrees with
this WKB expression within few percent for all molecules considered so far.

\subsection{Ammonia}
\label{sec-ammonia}

Equations (\ref{Eq:w_tun}), (\ref{Eq:Q_WKB}) show that sensitivity coefficient
logarithmically depends on the tunneling frequency. For example, for the
symmetric isotopologues of ammonia we get:
 \begin{subequations}
 \label{Q_nh3}
 \begin{align}
 \label{Q_nh3a}
 \mathrm{NH}_3:
 & \qquad\omega_{\rm tun} = 24\,\mathrm{GHz,}
 & Q_{\mu,\mathrm{tun}} = 4.5\,,
 \\
 \label{Q_nh3b}
 \mathrm{ND}_3:
 &\qquad  \omega_{\rm tun} = 1.6\,\mathrm{GHz,}
 & Q_{\mu,\mathrm{tun}} = 5.7\,.
 \end{align}
 \end{subequations}
Such a weak dependence on the tunneling frequency limits possible values of
the sensitivity coefficients for tunneling transitions in the microwave range:
$Q_{\mu,\mathrm{tun}}\lesssim 8$. This is quite good, compared to the
rotational sensitivity $Q_{\mu,\mathrm{rot}}\approx 1$, but smaller than the
best sensitivities in linear molecules considered above.

Let us consider mixed tunneling-rotational transitions, where tunneling goes
along with the change of the rotational quantum numbers. If we neglect
interaction between tunneling and rotational degrees of freedom we can write
approximate expressions for the frequency and the sensitivity of the mixed
inversion-rotational transition:
\begin{subequations}\label{appox_mix}
\begin{align}
 \label{w_mix}
 \omega_\mathrm{mix}
 &= \omega_r \pm \omega_\mathrm{tun}\,,\\
 \label{Q_mix}
 Q_{\mu,\mathrm{mix}}
 &= \frac{\omega_r}{\omega_\mathrm{mix}}
 \pm Q_{\mu,\mathrm{tun}}
 \frac{\omega_\mathrm{tun}}{\omega_\mathrm{mix}}\,.
\end{align}
\end{subequations}
We are particularly interested in the case when the minus sign in
\eqref{appox_mix} is realized and $\omega_\mathrm{mix}\ll\omega_\mathrm{tun}$.
For this case the tunneling sensitivity is enhanced by the factor
$\omega_\mathrm{tun}/\omega_\mathrm{mix}\gg 1$ and resultant sensitivity of
the mixed transition is inversely proportional to the transition frequency
$\omega_\mathrm{mix}$. Therefore, \textit{for the mixed transitions we can
have much higher sensitivities in the observable frequency range, then for the
purely tunneling transitions.}

Another important advantage of the mixed transitions is that there are usually
many of them each having different sensitivity. This means that we can have
very good control on possible systematics and reliably estimate the accuracy
of the results for $\mu$-variation.

The mixed transitions can not be observed in the symmetric isotopologues of
ammonia \eqref{Q_nh3}, but they are observed in the partly deuterated species
NH$_2$D and NHD$_2$. Unfortunately, for both of them the tunneling frequency
is much smaller then all rotational frequencies and sensitivities
\eqref{Q_mix} are not large~\cite{54}:
 \begin{align}\label{NH2D}
 \begin{array}{lll}
 \mathrm{NH}_2\mathrm{D}:
 &\qquad 0.10 \le Q_{\mu,\mathrm{mix}}\le 1.61\,,
 \\
 \mathrm{NHD}_2:
 &\qquad 0.27 \le Q_{\mu,\mathrm{mix}}\le 1.54\,.
 \end{array}
 \end{align}

Relatively small sensitivity coefficients for deuterated isotopologues of
ammonia \eqref{NH2D} and their low abundance does not allow to get strong
limits on $\mu$-variation, so we need to use tunneling ammonia line
\eqref{Q_nh3a}. It was observed from the several objects with the redshifts
about unity. Measuring radial velocities for rotational lines and for the
ammonia tunneling line we have $\Delta Q=3.5$ in \Eref{Sect2Eq8}, which is two
orders of magnitude larger than for optical lines. Because of that the ammonia
method allowed to place more stringent bounds on $\mu$-variation than bounds,
which follow from the optical spectra of the hydrogen molecule. However,
recent observations of the molecules with mixed tunneling-rotational
transitions provide even higher sensitivity to $\mu$-variation.

\subsection{Mixed tunneling-rotational transitions and effective Hamiltonians}
\label{sect-mgk5}

Equations \eqref{appox_mix} show that high sensitivity mixed transitions are
possible when tunneling frequency is of the same order of magnitude as
rotational constants. However, in this case tunneling and rotational degrees
of freedom start to interact and the accuracy of approximation
\eqref{appox_mix} decreases. A much better approximation can be reached with
the help of the effective Hamiltonians, which describe rotational and
tunneling degrees of freedom and their interactions with each other. At
present the state of the art effective Hamiltonians can include on the order
of hundred parameters. These parameters are fitted to the experimentally known
transitions and provide an accuracy on the ppm scale, or better.

%----------------------- Figure 6
\begin{figure*}[t]
 \includegraphics[scale=0.60]{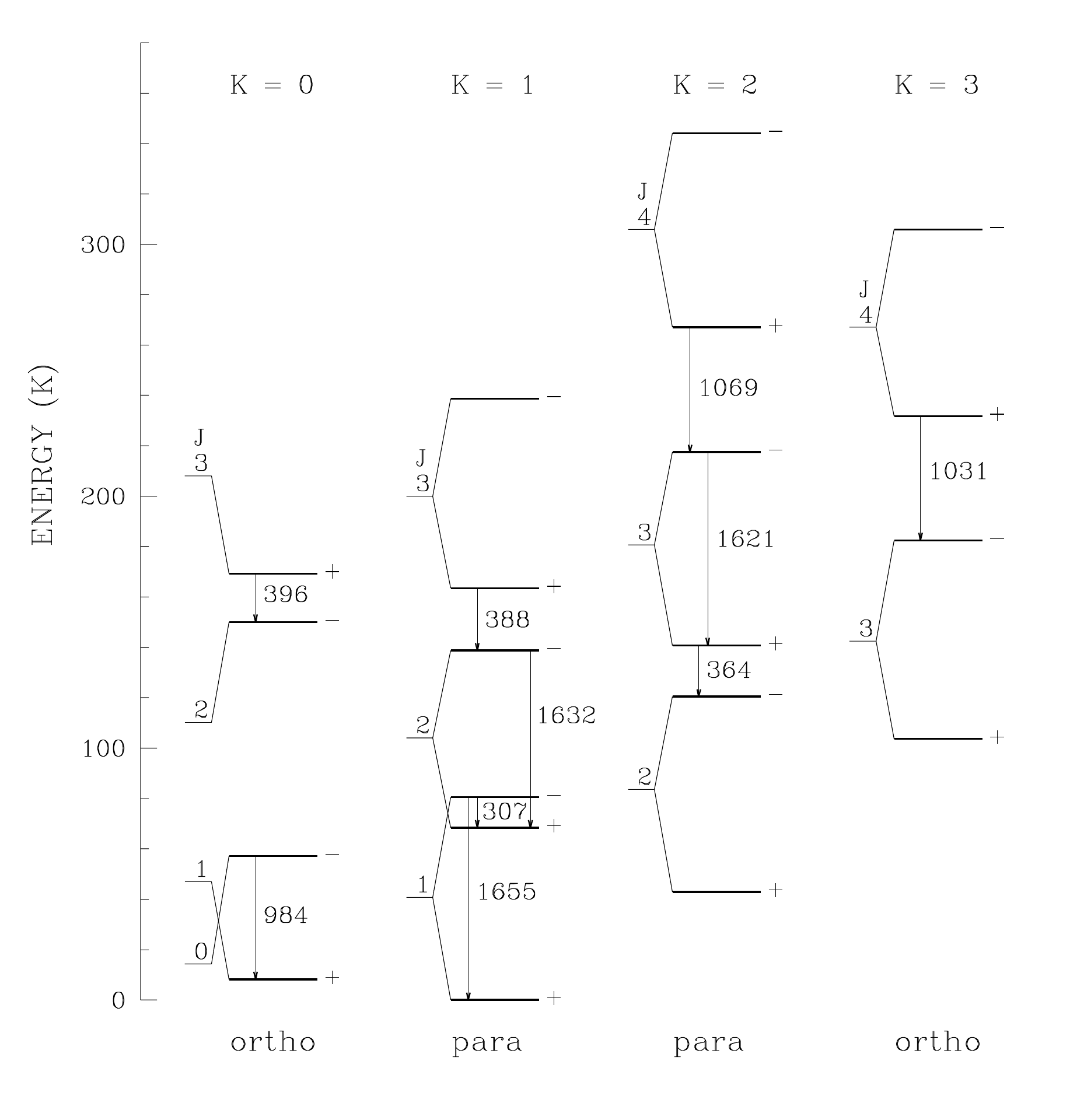}
 \caption{Tunneling-rotational spectrum of H$_3$O$^+$ molecule. Several
 low-frequency tunneling and mixed transitions are marked with vertical
 arrows. Their frequencies are shown in GHz.}
\label{fg6}
\end{figure*}

When such Hamiltonians are used to find sensitivity coefficients $Q_\mu$ we
need to know how all the parameters depend on $\mu$.
It was shown in~\cite{60} that this can usually be done only
within an accuracy of a few percent. The final accuracy for the large
$Q$-factors is somewhat lower because of the instability of \Eref{Q_mix}.
Because of that we need not complex effective Hamiltonians but their
simplified versions with considerably smaller numbers of fitting parameters
can be used instead.

\subsection{Hydronium and peroxide}
\label{h3o+_and_h2o2}

Let us start with hydronium molecule H$_3$O$^+$~\cite{52}. This
molecule is a symmetric top. It is similar to ammonia, but flatter. Tunneling
frequency is almost 50 times larger and comparable to rotational intervals.
The tunneling umbrella mode does not change the symmetry and does not
contribute to the angular momentum of the molecule. Because of that the
tunneling-rotational interaction is reduced to the centrifugal corrections to
the tunneling frequency~\cite{105}. 

The tunneling-rotational spectrum of hydronium is shown in \fref{fg6}. It
consists of the $J$ ladders for each quantum number $K$, where $K$ is
projection of the angular momentum on the molecular axis. Due to the tunneling
each rotational level is split in two states with different parity $p$. For
$K=0$ the permutation symmetry of the hydrogen nuclei allows only one of these
levels, while for $K>0$ both levels are present.

%-----------------------------Table VII
\begin{table*}[t!]
\caption{Sensitivities of the low frequency mixed inversion-rotational
transitions in hydronium H$_3$O$^+$.
Molecular states are labeled with quantum numbers $J_K^p$.
}
\label{tab-7}

\begin{tabular}{ccrrr}
\hline\hline
 \multicolumn{2}{c}{Transition}
 &\multicolumn{2}{c}{Frequency (MHz)}
 &\multicolumn{1}{c}{$Q_\mu$}\\
 \multicolumn{1}{c}{Upper}
 &\multicolumn{1}{c}{Lower}
 &\multicolumn{1}{c}{Theory}
 &\multicolumn{1}{c}{Exper.}\\
 \hline\\[-3mm]
 $ 1_1^- $&$ 2_1^+ $&  307072 &   307192.4 &$  6.4(5) $\\
 $ 3_2^+ $&$ 2_2^- $&  365046 &   364797.4 &$ -3.5(5) $\\
 $ 3_1^+ $&$ 2_1^- $&  389160 &   388458.6 &$ -3.1(4) $\\
 $ 3_0^+ $&$ 2_0^- $&  397198 &   396272.4 &$ -3.0(4) $\\
 $ 0_0^- $&$ 1_0^+ $&  984690 &   984711.9 &$  2.7(2) $\\
 $ 4_3^+ $&$ 3_3^- $& 1031664 &  1031293.7 &$ -0.6(2) $\\
 $ 4_2^+ $&$ 3_2^- $& 1071154 &  1069826.6 &$ -0.5(2) $\\
 $ 3_2^- $&$ 3_2^+ $& 1621326 &  1621739.0 &$  2.0(1) $\\
 $ 2_1^- $&$ 2_1^+ $& 1631880 &  1632091.0 &$  2.0(1) $\\
 $ 1_1^- $&$ 1_1^+ $& 1655832 &  1655833.9 &$  2.0(1) $\\[1mm]
\hline\hline
\end{tabular}
\end{table*}

In \fref{fg6} we see four mixed transitions with frequencies around 300
GHz, which is few times smaller than the tunneling frequency that is about
1.6 THz. \tref{tab-7} shows that these transitions have enhanced sensitivity
to $\mu$ variation ($Q_{\mu,\mathrm{tun}}= 2.0\pm0.1$). Those transitions,
whose frequencies decrease when tunneling frequency increases have negative
sensitivity coefficients $Q_\mu$. We conclude that hydronium has several mixed
transitions with sensitivities of both signs and the maximum $\Delta Q_\mu$ is
around 10. Other isotopologues of hydronium have even higher sensitivities
\cite{53}, but up to now they have not been observed in the interstellar medium.

Another molecule where tunneling frequency is comparable to rotational
constants, but tunneling-rotational interaction is rather weak, is peroxide
H$_2$O$_2$~\cite{38,67}. In equilibrium geometry
H$_2$O$_2$ is not flat; the angle 2$\gamma$ between two HOO planes is close to
113\textdegree. Two flat configurations correspond to local maxima of
potential energy; the potential barrier for \textit{trans} configuration
($2\gamma=\pi$) is significantly lower, than for \textit{cis} configuration
($\gamma=0$): $U_\pi\approx 400$ \cmo\ and $U_0\approx 2500$ \cmo. To a first
approximation one can neglect the tunneling through the higher barrier. In
this model peroxide is described by a slightly asymmetric oblate top with
inversion tunneling mode, similar to ammonia and hydronium.

%-------------------------Table VIII
\begin{table}[tbh]
\caption{Numerical calculation of the $Q$-factors for low frequency mixed
transitions in peroxide H$_2$O$_2$ using effective Hamiltonian. Experimental
frequencies are taken from JPL Catalogue~\cite{84}. 
$E_\mathrm{up}$ is upper state energy in Kelvin.} \label{tab-8}
\begin{tabular}{c@{ -- }ccrdd}
\hline\hline
 \multicolumn{2}{c}{$J_{K_A,K_C}(\tau)$}
 &\multicolumn{1}{c}{$E_\mathrm{up}$}
 &\multicolumn{2}{c}{$\omega$ (MHz)}
 &\multicolumn{1}{c}{$Q_\mu$}
\\
 \cline{1-2}\cline{4-5}
 \multicolumn{1}{c}{upper}
 &\multicolumn{1}{c}{lower}
 &\multicolumn{1}{c}{(K)}
 &\multicolumn{1}{c}{theory}
 &\multicolumn{1}{c}{exper.}
 &
\\
\hline
\multicolumn{6}{c}{Transitions below 100 GHz}\\
$ 0_{0,0}(3) $&$ 1_{1,0}(1) $& 17& 14818.8    &    14829.1  &   +36.5(2.9) \\
$ 2_{1,1}(1) $&$ 1_{0,1}(3) $& 21& 37537.0    &    37518.28 &   -13.0(1.2) \\
$ 1_{0,1}(3) $&$ 1_{1,1}(1) $& 19& 67234.5    &    67245.7  &    +8.8(6) \\
$ 2_{0,2}(3) $&$ 2_{1,2}(1) $& 24& 68365.3    &    68385.0  &    +8.7(6) \\
$ 3_{0,3}(3) $&$ 3_{1,3}(1) $& 31& 70057.4    &    70090.2  &    +8.5(6) \\
$ 4_{0,4}(3) $&$ 4_{1,4}(1) $& 41& 72306.0    &    72356.4  &    +8.3(6) \\
$ 5_{0,5}(3) $&$ 5_{1,5}(1) $& 53& 75104.6    &    75177.4  &    +8.0(6) \\
$ 6_{0,6}(3) $&$ 6_{1,6}(1) $& 68& 78444.7    &    78545.4  &    +7.7(6) \\
$ 3_{1,2}(1) $&$ 2_{0,2}(3) $& 28& 90399.8    &    90365.51 &    -4.8(5) \\[1mm]
\multicolumn{6}{c}{Transitions observed from ISM in~\cite{7}}\\
$ 3_{0,3}(3) $&$ 2_{1,1}(1) $& 31&219163.2    &   219166.9  &    +3.4(2) \\
$ 6_{1,5}(1) $&$ 5_{0,5}(3) $& 66&252063.6    &   251914.68 &    -1.1(2) \\
$ 4_{0,4}(3) $&$ 3_{1,2}(1) $& 41&268963.7    &   268961.2  &    +3.0(2) \\
$ 5_{0,5}(3) $&$ 4_{1,3}(1) $& 53&318237.7    &   318222.5  &    +2.7(1) \\[1mm]
\hline\hline
\end{tabular}
\end{table}

The sensitivity coefficients for the mixed transitions in peroxide were
calculated in~\cite{50}. Results of these calculations are shown in
\tref{tab-8}. Molecular states are labeled with the rotational quantum
numbers $J$, $K_A$, and $K_C$ and the tunneling quantum number $\tau$~\cite{38}. 
Transitions with the frequencies below 100 GHz were found to have
rather high sensitivities of both signs. Several transitions of peroxide were
recently observed from interstellar medium (ISM) in~\cite{7}. 
These transitions have higher frequencies and smaller sensitivities to
$\mu$-variation. Nevertheless, even for these transitions the maximum value of
$\Delta Q_\mu$ is about 4.5.

\subsection{Molecules with hindered rotation: methanol and methylamine}
\label{sect-mgk6}

Hindered rotation is one of the examples of the large amplitude internal
motions in non rigid molecules. In the discussion of the peroxide molecule in
the previous subsection, we neglected the tunneling through the higher
\textit{cis} barrier. For the excited vibrational states tunneling through
both barriers can take place leading to the hindered rotation of one HO group
in respect to another. Many molecules which include CH$_3$ group have three
equivalent minima at 120\textdegree\ to each other. Hindered rotation in such
molecules can take place already for the ground vibrational state. When the
tunneling frequencies are comparable to the rotational ones, such molecules
have very rich microwave spectra with a large number of mixed transitions.
Another distinctive feature of these molecules is strong interaction between
the internal (hindered) and overall rotations. One of the simplest molecules
of this type is methanol CH$_3$OH.

The basic theory of the non-rigid tops with internal rotation was established
in the 1950s~\cite{65,37} and the main features of
the methanol spectrum were explained. Later on the theory was refined many
times and currently there is a very impressive agreement between the theory
and experiment~\cite{3,76,103,48}.

%---------------------- Figure 7
\begin{figure*}[tbh]
 \includegraphics[scale=1.0]{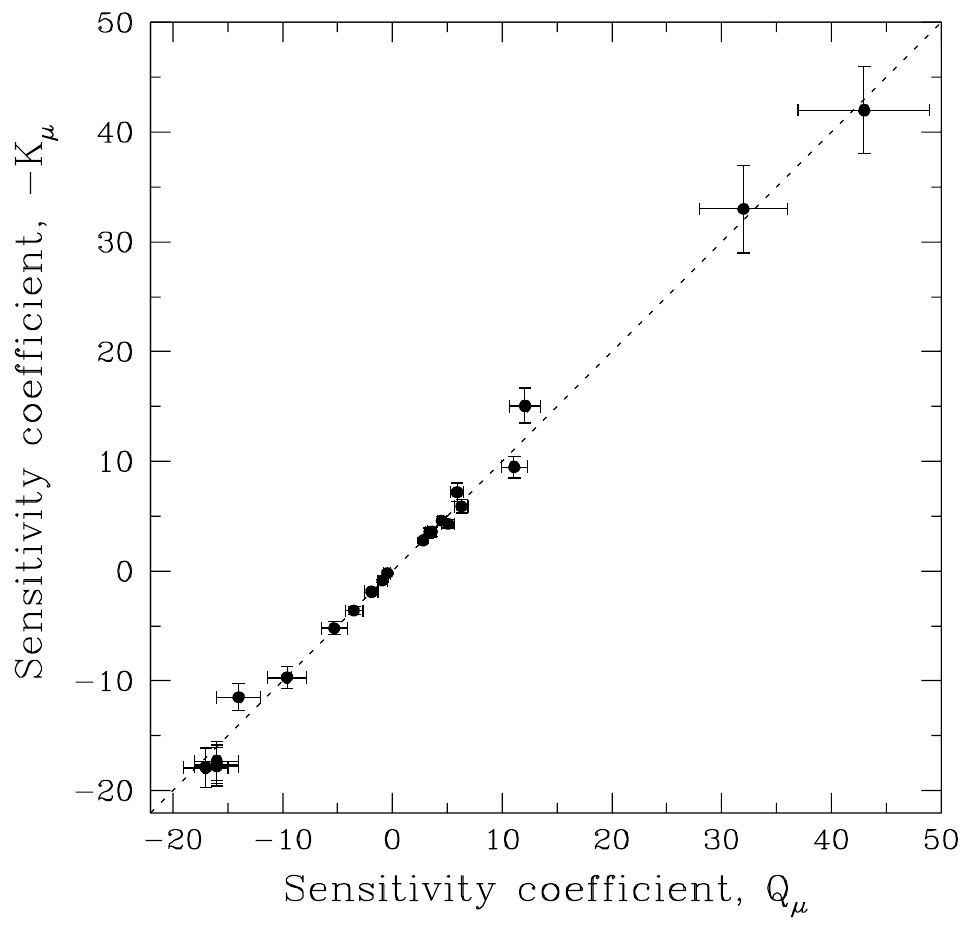}
 \caption{Comparison of the sensitivity coefficients for CH$_3$OH from
\cite{43} and \cite{60}. The former used the sensitivity coefficients
 $K_\mu$ defined as $K_\mu=-Q_\mu$. This corresponds to the different
 definition of the mass ratio: $m_{\rm p}/m_{\rm e}$ instead of $m_{\rm e}/m_{\rm p}$,
 which is used in the present review.}
\label{fg7}
\end{figure*}

The sensitivity coefficients to the $\mu$-variation for methanol microwave
transitions were calculated independently in~\cite{43,44} 
and in \cite{60}. The first group used the state of the art effective
Hamiltonian~\cite{103}, which included 120 fitting parameters. The
second group used a much simpler model~\cite{86}. The rotational
part $H_\mathrm{rot}$ was that of the slightly asymmetric top and included the
rotational constants $A$, $B$, and $C$ ($A \approx B$). The hindered rotation was
described by the Hamiltonian
\begin{equation}\label{H1}
    H_\mathrm{hr}=-F\frac{\mathrm{d}^2}{\mathrm{d}\omega^2}
    + \frac{V_3}{2}\left(1-\cos\,3\omega \right)\,,
\end{equation}
where the kinetic coefficient $F$ was proportional to $\mu$ and the electronic
potential $V_3$ was independent on $\mu$. The angle $\omega$ described
position of the OH group in respect to the CH$_3$ top. This model did not
include centrifugal distortions. The interaction of the internal rotation with
the overall rotation was described by a single parameter $D$, which scaled
linearly with $\mu$~\cite{65}. Altogether this model had 6 parameters.

%----------------------- Figure 8
\begin{figure*}[t]
 \includegraphics[scale=0.50]{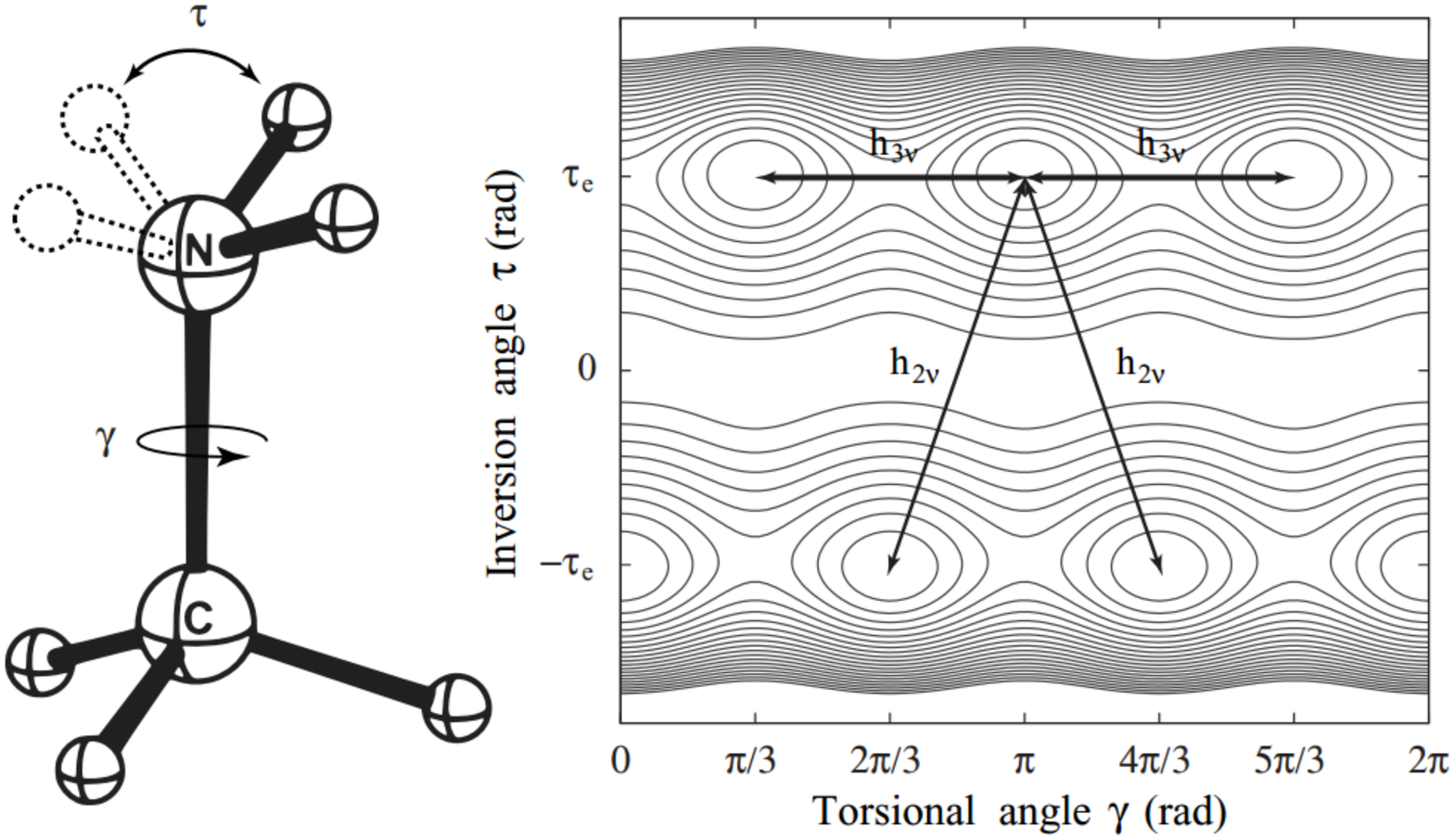}
 \caption{Schematic representation of methylamine and variation
of the potential energy of methylamine as function of the relative rotation
$\gamma$ of the CH$_3$ group with respect to the amine group about the CN bond
and the angle $\tau$ of the two hydrogen atoms of the NH$_2$ group with
respect to the CN bond. The two large amplitude motions, corresponding to
inversion $h_{2v}$ and hindered rotation $h_{3v}$ are schematically indicated
by the arrows. Note that inversion of the NH$_2$ group is accompanied by a
$\pi/3$ rotation about the CN bond of the CH$_3$ group with respect to the
amine group.}
 \label{fg8}
\end{figure*}

Both effective Hamiltonians were diagonalized for several sets of parameters,
which correspond to an increased and decreased $\mu$ and the sensitivity
coefficients were found by the numerical differentiation. The comparison of
the two calculations is given in \fref{fg7}. We see that in spite of a
significant difference in complexity of the models the results are in good
agreement and of the comparable accuracy. As we discussed above, the latter is
mostly determined by ambiguity in the $\mu$-scaling of model parameters.

The sensitivity coefficients for the mixed transitions in methanol span from $-17$ to
+43, which corresponds to $|\Delta Q_\mu| \sim 60$. This is more than an order
of magnitude larger than in ammonia method. Moreover, in methanol we have
a large number of strong lines with different sensitivities and can effectively
control possible systematic effects. Until very recently methanol was observed
only at small redshifts, but in 2011 it was first detected in the microwave
survey towards the object PKS 1830-211 at redshift $z=0.89$~\cite{75}. 
This means that at present methanol can be used as a very sensitive tool to probe
$\mu$-variation on a cosmological timescale~\cite{4,26}.

In the same survey~\cite{75}, a large number of rather complex molecules
were detected for the first time at high redshift. In particular, the list
includes methylamine~--- yet another molecule with tunneling motion. In
contrast to all previously discussed molecules, methylamine has two tunneling
modes. First is hindered rotation of the NH$_2$ around CH$_3$ top, which is
similar to that in methanol. Second is a wagging mode when the NH$_2$ group
flips over to the other side (see \fref{fg8}). Both modes contribute to
the angular momentum of the molecule and, therefore, strongly interact with
the overall rotation.

The spectrum of methylamine is also very rich. The effective Hamiltonian must
include both tunneling motions and their interactions with each other and with
the overall rotation. Therefore, even the simplest form of this Hamiltonian is
quite complex and we will not discuss it here. Calculations of the sensitivity
coefficients were recently done in~\cite{42}. It was found that
they lie in the range $-24\le Q_\mu\le 19$. However, the lines, which were
observed in~\cite{75} at $z=0.89$ have sensitivities close to 1. Up
to now neither of the more sensitive lines of methylamine has been observed at
high redshifts.

There are several other molecules with mixed tunneling-rotational spectra,
for example N$_2$H$_4$ and CH$_3$SH. The former has three
tunneling modes which strongly interact with rotation. Thus, we should expect 
very complex spectrum. This molecule is predicted to form in Jupiter's and
Titan's atmospheres~\cite{74,94}. The latter is
similar to methanol and exhibits hindered rotation and complex spectrum~\cite{9}.
Effective Hamiltonians for many of these molecules 
are known, but no other calculations of the sensitivity coefficients have 
been done so far (preliminary results for CH$_3$SH show that there are 
transitions with high sensitivities of both signs). If any new sufficiently 
low frequency mixed transitions are observed from the interstellar medium, 
it is possible to calculate respective sensitivity coefficients using the 
methods outlined in this section.

\section{Summary and conclusions}
\label{sect-6}

As we discussed in the previous sections the constraints on the possible
variation of fundamental constants are an efficient method of testing the
equivalence principle which is a basic assumption of General Relativity. These
constraints can be derived from a wide variety of atomic and molecular
transitions observed in laboratory, solar and extra solar systems, and at very
early cosmological epochs up to a redshift of order $z \sim 5-6$ where
molecular and atomic transitions have been recently detected and observed with
a sufficiently high spectral resolution~\cite{59,58}. 
Radio astronomical observations of the NH$_3$ molecule in two
distant galaxies provide tight constraints at the \dmm\ $< 1\times10^{-6}$
level at $z = 0.89$~\cite{36} and $z = 0.69$~\cite{46}. Even
deeper bounds were deduced from observations of the CH$_3$OH molecule in the
$z = 0.89$ galaxy: \dmm\ $< 3\times10^{-7}$~\cite{26}, and \dmm\
$< 1\times10^{-7}$~\cite{4}. 

To probe $\alpha$ and $\mu$ at the level of $10^{-8}$ or $10^{-9}$, at least
two main requirements should be fulfilled: ($i$) increasing precision of the
laboratory measurements of the rest frame frequencies of the most sensitive
molecular transitions discussed in this review, and ($ii$) increasing
sensitivity and spectral resolution of astronomical observations.

The most promising molecular transitions are those of a mixed nature,
where there are two, or more, competing contributions to the transition
energy. We get strong enhancement of the sensitivity to the variation of the
fundamental constants when the resultant transition frequency is much smaller
than individual contributions. This happens, for example, for some mixed
tunneling-rotational transitions. Diatomic radicals give another example,
where spin-orbit interaction is competing with Coriolis interaction. As a
result we have strong enhancement of the sensitivity coefficients for the
$\Lambda$-doublet transitions. There are other known examples, which are more
relevant for the laboratory experiments~\cite{16,8}. 
It is possible that more examples will be found both for the
laboratory and astrophysical studies.
The methods described in this review allow us to calculate sensitivity
coefficients for any microwave and submillimeter molecular transitions of
interest.

\end{document}